\documentclass[12pt]{article}
\usepackage{amsmath,amssymb}
\usepackage{bm}
\usepackage{graphicx,color}
\usepackage{wrapfig}
\begin{document}
\title{
\begin{flushright}
\ \\*[-80pt] 
\begin{minipage}{0.2\linewidth}
\normalsize
%arXiv:YYMM.NNNN \\
\end{minipage}
\end{flushright}
{\Large \bf   Probing the  high scale SUSY in  CP violations  \\ of $K$, $B^0$ and $B_s$ mesons
\\*[20pt]}}
\author{
\centerline{Morimitsu~Tanimoto\footnote{E-mail address: tanimoto@muse.sc.niigata-u.ac.jp} \ \ and \ \
Kei~Yamamoto\footnote{E-mail address: yamamoto@muse.sc.niigata-u.ac.jp}}
\\*[20pt]
\centerline{
\begin{minipage}{\linewidth}
\begin{center}
{\it \normalsize
Department of Physics, Niigata University,~Niigata 950-2181, Japan }
\end{center}
\end{minipage}}
\\*[70pt]}

\date{
\centerline{\small \bf Abstract}
\begin{minipage}{0.9\linewidth}
\vskip  1 cm
\small
We probe the high scale SUSY at  $10-50$ TeV in the CP violations of $K$, $B^0$ and $B_s$ mesons.
 In order to estimate the contribution of the squark flavor mixing to these CP violations,
we discuss the squark mass spectrum, which is consistent with
   the recent Higgs discovery. Taking  the universal soft parameters at the SUSY breaking scale, we obtain
   the squark mass spectrum at $10$ TeV and $50$ TeV,  where
 the SM emerges. Then, the $6\times 6$ mixing matrix between  down-squarks and down-quarks is discussed 
 by input of the experimental data of $K$, $B^0$ and $B_s$ mesons.
  It is found that   $\epsilon_K$ is most  sensitive to the high scale SUSY.  
 The SUSY contributions for the time-dependent CP asymmetries   $S_{ J/\psi K_S}$ and 
 $S_{ J/\psi \phi}$ are  $6-8 \%$ at the SUSY scale of $10$ TeV.
We also discuss the  SUSY contribution to the chromo-EDM of the strange quark.
 \end{minipage}
}

\begin{titlepage}
\maketitle
\thispagestyle{empty}
\end{titlepage}

\section{Introduction}
\label{sec:Intro}
 Although the supersymmetry (SUSY) is one of the most attractive candidates for the new physics,  the SUSY signals have not been observed yet.
 Therefore, the recent searches for new particle at the LHC give us important constraints for SUSY.
Since the lower bounds of the superparticle masses increase gradually, 
the squark and the gluino masses are supposed  
to be at the higher scale than $1$ TeV ~\cite{squarkmass}. 
Moreover, the SUSY model has been seriously constrained  by the Higgs discovery, in which
the Higgs mass is  $126$ GeV~\cite{Higgs}. 

These facts suggest a class of SUSY models with heavy sfermions.
If  the SUSY is broken with the breaking scale $10-100$ TeV,  the squark and slepton masses
are expected to be  also ${\cal O}(10-100)$ TeV.
Then, the lightest Higgs mass can be pushed up to $126$ GeV, while all SUSY particle  can be out of the reach of the LHC experiment.
Therefore, the indirect search of the SUSY particles becomes important in the low energy
flavor physics \cite{Altmannshofer:2013lfa,Moroi:2013sfa}.

The flavor physics is on the new stage in the light of LHCb data.
The LHCb collaboration has reported 
new data of the CP violation of the $B_s$ meson and the branching ratios 
of rare $B_s$ decays~\cite{Bediaga:2012py}-\cite{LHCb:2011ab}.
For many years the CP violation in the $K$ and $B^0$ mesons 
has been successfully understood within the framework of the standard model (SM), 
so called Kobayashi-Maskawa (KM) model \cite{Kobayashi:1973fv}, 
where the source of the CP violation is the KM phase 
in the quark sector with three families. 
However, the new physics has been expected to be indirectly discovered
in the precise data of $B^0$ and $B_s$ meson decays at the LHCb experiment and the further coming  experiment, Belle II. 

While, there are new sources of the CP violation if the SM is 
extended to the SUSY models. The soft squark mass matrices contain 
the CP violating phases, which contribute to the flavor changing 
neutral current (FCNC) with the CP violation \cite{Gabbiani:1996hi}. 
We can expect the SUSY effect in the CP violating phenomena. 
However, the clear deviation from the  SM prediction
has not been observed yet in the LHCb experiment~\cite{Bediaga:2012py}-\cite{LHCb:2011ab}.
Therefore, we should carefully study the CP-violation phenomena.

The LHCb collaboration presented the time dependent CP asymmetry 
in the non-leptonic $B_s\to {J/\psi \phi}$ decay \cite{Aaij:2013oba,LHCb:2011aa,LHCb:2011ab}, which gives a constraint of 
the SUSY contribution on the $b\to s$ transition. 
%They have  also reported the first measurement of the CP violating phase 
%in the $B_s \to \phi \phi$ decay \cite{Aaij:2013qha}. 
%If there is the contribution of the squark flavor mixing in the FCNC, we expect to observe 
%the sizeable time dependent CP asymmetry in this process because the SM contribution
%is tiny.
%This decay process is occurred at the one-loop level in the SM, where the CP violating phase
%is very small. On the other hand, the gluino-squark mediated flavor changing process 
%provides new CP violating phases. Thus, the CP asymmetry of $B_s \to \phi \phi$
%is expected to be deviated considerably from the SM one.
In this work, we discuss the sensitivity of the high scale SUSY contribution to the CP violation of $K^0$, $B_d$ and $B_s$ mesons.
For these decay modes,  the most important process of the SUSY contribution  is
 the gluino-squark mediated  flavor changing process \cite{King:2010np}- \cite{Hayakawa:2013dxa}.
This FCNC effect is constrained by the CP violations in  $B^0\to {J/\psi K_S}$ and
 $B_s\to {J/\psi \phi}$ decays. 
 The CP violation of $K$ meson, $\epsilon_K$,  also provides a severe constraint to the gluino-squark mediated FCNC. In the SM, $\epsilon_K$ is proportional to 
 $\sin (2\phi_1)$ which is derived from the time dependent CP asymmetry 
 in $B^0\to J/\psi K_s$  decay
\cite{Buras:2008nn}. 
 The relation between $\epsilon_K$ and  $\sin (2\phi_1)$ is examined 
 by taking account of  the gluino-squark mediated FCNC \cite{Mescia:2012fg}.
 
 %The decay processes with the $b\to s$ transition also give the constraints of 
 %the SUSY contribution.
 % The typical constraint comes from  the $B^0\to X_s\gamma $ decay, 
%and  the time dependent CP asymmetries 
%in $B^0\to \phi K_S$ and $B^0\to \eta ' K^0$ decays~\cite{PDG,Amhis:2012bh}. 

The time dependent CP asymmetry of 
$B^0\to \phi K_S$ and  $B^0\to \eta ' K^0$ decays
are also  considered as  typical processes  to search for the gluino-squark mediated FCNC 
because the penguin amplitude dominates this process.
Furthermore, we  discuss the semileptonic CP asymmetries of $B^0$ and $B_s$ mesons,
which can probe the SUSY contribution.

In addition, it is remarked  that the upper-bound of the chromo-EDM(cEDM) of the strange quark gives a severe constraint for the gluino-squark mediated $b\to s$ transition
\cite{Hisano:2003iw}-\cite{Fuyuto:2012yf}.
The recent work shows us that the cEDM is sensitive to the high scale SUSY \cite{Fuyuto:2013gla}.

  %%%%%%%%%%%%%%%%%%%%%%%%%%%%%%%%%%%%%%%

  In order to estimate the gluino-squark mediated FCNC 
  of the  $K$, $B^0$ and $B_s$ meson   for arbitrary squark mass spectra,
  we work in the basis of  the squark mass eigenstate. 
  %Then,  we should input the squark spectrum  to predict the magnitude of the FCNC.
 There are three reasons why the SUSY contribution to the FCNC considerably depends on the squark mass spectrum.
  The first one is that the GIM mechanism works in the squark flavor mixing, and 
  the second one is that the loop functions depend on the mass ratio of squark and gluino.
  The last one is that we need the mixing angle between the left-handed sbottom and right-handed sbottom,
  which dominates the $\Delta B=1$ decay processes.
   Therefore, we discuss the squark mass spectrum, which is consistent with
   the recent Higgs discovery.
   Taking  the universal soft parameters at SUSY breaking scale, we obtain
   the squark mass spectrum at the matching scale where
 the SM emerges, by using the Renormalization Group Equations (REG's) of the soft masses.
  Then, the $6\times 6$ mixing matrix between  down-squarks and down-quarks is examined 
 by input of the experimental data.

%In order to obtain reliable predictions, %%%%%%%%%%
%%%%%%%%%%%%%%%%%%%%%%%%%%%%%
In section 2, we discuss the squark and gluino spectra.
In section 3, we present the formulation of the  CP violation in terms of the squark flavor mixing, and 
we present our numerical results in section 4.
Section 5 is devoted to the summary. 
Relevant formulations are presented in Appendices A, B and C.

%%%%%%%%%%%%%%%%%%%%%%%%%%%%%%%%%%%%%%%%%%%%%%%%%%%%%%%%%%%%%%%%%%%%%%%%%%%%%%%%%
%%%%%%%%%%%%%%%%   Squark flavor mixing in CP violation of $B$ mesons  %%%%%%%%%%
%%%%%%%%%%%%%%%%%%%%%%%%%%%%%%%%%%%%%%%%%%%%%%%%%%%%%%%%%%%%%%%%%%%%%%%%%%%%%%%%%

\section{ SUSY Spectrum}
\label{sec:spectrum}
We consider the SUSY model with heavy sfermions.
If the squark and slepton masses
are expected to be  also ${\cal O}(10)$ TeV, the lightest Higgs mass can be pushed up to $126$ GeV.

Let us obtain the SUSY particle mass spectrum in the framework of
the minimal supersymmetric standard model (MSSM), which is consistent with the observed Higgs mass.
The numerical analyses have been given in refs.
\cite{Delgado:2013gza, Giudice:2006sn}.
At the SUSY breaking scale $\Lambda$, the quadratic terms in the MSSM potential is given as
\begin{equation}
 V_2=m_1^2 |H_1|^2+ m_2^2 |H_2|^2+m_3^2 (H_1\cdot H_2+ h.c.) \ ,
\end{equation}
where we define $m_1^2=m_{H_1}^2+|\mu|^2$ and  $m_2^2=m_{H_2}^2+|\mu|^2$ in terms of 
the soft breaking mass  $m_{H_i}$ and the supersymmetric Higgsino mass $\mu$.
The mass eigenvalues at  the $H_1$ and $\tilde H_2\equiv \epsilon H_2^*$ system are given
\begin{equation}
 m_\mp^2 = \frac{m_1^2+m_2^2}{2}\mp \sqrt{\left (\frac{m_1^2-m_2^2}{2}\right )^2+m_3^4} \ .
\end{equation}
Suppose that the MSSM matches with the SM at the SUSY mass scale $Q_0\equiv m_0$.
Then, the smaller one $m_{-}^2$ is identified to be the mass squared of the SM Higgs $H$ with the tachyonic mass.
On the other hand, the larger one $m_{+}^2$ is  the mass squared of the orthogonal combination ${\cal H}$, which
is decoupled from the SM at $Q_0$, that is, $m_{{\cal H}}\simeq Q_0$
. Therefore, we have
\begin{eqnarray}
 m_{-}^2=-m^2(Q_0)\ , \qquad  m_{+}^2=m_{\cal H}^2(Q_0) =m_1^2+m_2^2+m^2 \ ,
\end{eqnarray}
with 
\begin{eqnarray}
 m_{3}^4=(m_1^2+m^2)(m_2^2+m^2) \ ,
\end{eqnarray}
which lead to the mixing angle between $H_1$ and $\tilde H_2$, $\beta$ as
\begin{eqnarray}
 \tan^2\beta=\frac{m_1^2+m^2}{m_2^2+m^2} \ ,
\end{eqnarray}
where
\begin{eqnarray}
 H=  \cos\beta H_1 +\sin\beta \tilde H_2  \ , \nonumber \\
 {\cal H} =-\sin\beta H_1 +\cos\beta \tilde H_2  \ .
\end{eqnarray}
Thus,  the Higgs mass parameter $m^2$ is expressed in terms of $m_1^2$, $m_2^2$ and $\tan\beta$:
\begin{eqnarray}
m^2=\frac{m_1^2-m_2^2\tan^2\beta}{\tan^2\beta -1} \ .
\end{eqnarray}
Below the energy scale $Q_0$, in which the SM emerges, the scalar potential is just the SM one as follows:
\begin{equation}
 V_{SM}=-m^2 |H|^2+\frac{\lambda_H}{2} |H|^4 \ .
\end{equation}
Here, the Higgs coupling $\lambda_H$ is given in terms of the SUSY parameters as
 \begin{equation}
 \lambda_H(Q_0)=\frac{1}{4} (g^2+g'^2)\cos^2 2\beta +\frac{3h_t^2}{8\pi^2} X_t^2 \left (1- \frac{X_t^2}{12}\right ) \ ,
\end{equation}
where
 \begin{equation}
 X_t=\frac{A_t(Q_0)-\mu(Q_0)\cot\beta}{ Q_0 } \ ,
\end{equation}
and $h_t$ is the top Yukawa coupling of the SM.
 The parameters $m_2$ and $\lambda_H$ run with the SM Renormalization Group Equation (RGE) down to the electroweak scale $Q_{EW}=m_H$, and then give
\begin{equation}
 m_H^2=2m^2(m_H)=\lambda_H(m_H)v^2\ .
\end{equation}
It is easily seen  that the VEV of Higgs, $\langle H \rangle $ is $v$,  and $\langle {\cal H}\rangle=0$,
taking account of $\langle H_1\rangle =v\cos\beta$ and $\langle H_2\rangle =v\sin\beta$,
 where $v=246$GeV.
 
travel 

Let us  fix $m_H=126$GeV, which gives $\lambda_H(Q_0)$ and $m^2(Q_0)$. This experimental input constrains the
SUSY mass spectrum of MSSM.
We consider the some universal soft breaking parameters at the SUSY breaking scale $\Lambda$ as follows:
\begin{eqnarray}
&&m^2_{\tilde Q_i}(\Lambda)=m^2_{\tilde U^c_i}(\Lambda)=m^2_{\tilde D^c_i}(\Lambda)=m_0^2 \ (i=1,2,3) \ , \nonumber \\
&&M_1(\Lambda)=M_2(\Lambda)=M_3(\Lambda)=m_{1/2},  \ , \nonumber \\
&&m_{H_1}^2({\Lambda})= m_{H_2}^2({\Lambda})=m_0^2 \ , \nonumber \\
&&A_U({\Lambda})=A_0 y_U(\Lambda), \quad A_D({\Lambda})=A_0 y_D(\Lambda), \quad A_E({\Lambda})=A_0 y_E(\Lambda).
\label{universal}
\end{eqnarray}
Then,  there is no flavor mixing at this scale if the universal soft masses are exactly satisfied.
Different RGE effects for each flavor evolve the squark flavor mixing 
  at the lower energy scale, which is controlled by the CKM mixing matrix.
 Since we take squark flavor mixing  as free parameters  at the low energy, 
 this universality condition has to be considered as an approximation and non-vanishing off diagonal squark mass matrix elements are  introduced at the $\Lambda$ scale.
We will show typical magnitudes of those off-diagonal elements in the numerical result
 to understand the level of our approximation.in the numerical result.

Now, we have the SUSY five parameters, $\Lambda$, $\tan\beta$, $m_0$, $m_{1/2}$, $A_0$,
where $Q_0=m_0$. In addition to these parameters, we take $\mu=Q_0$.
Inputing  $m_H=126$GeV and taking $m_{{\cal H}}\simeq Q_0$, we can obtain the SUSY spectrum
 for the fixed $Q_0$ and $\tan\beta$. 

\begin{figure}[h]
%\begin{minipage}[]{0.45\linewidth}
\begin{center}
\includegraphics[width=12cm]{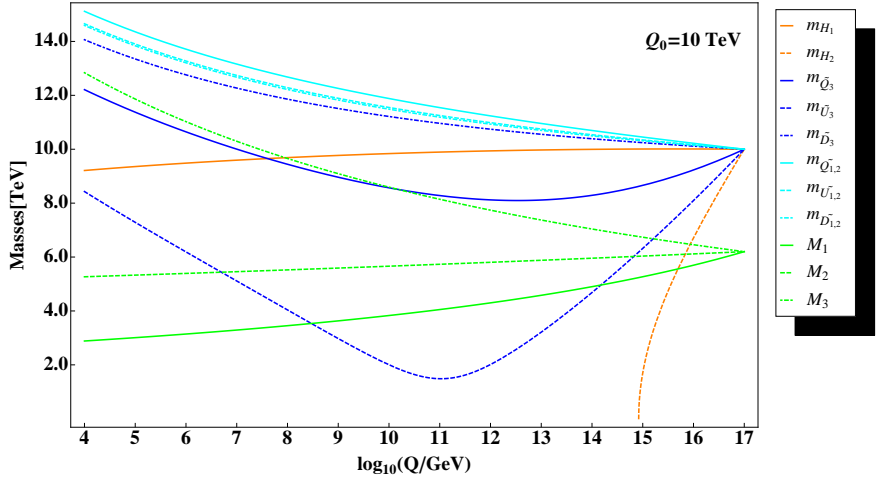}
\hspace{1cm}
%\includegraphics[width=7cm]{fig1b.eps}
%\hspace{0.1cm}
\caption{Running of SUSY mass parameters from $\Lambda=10^{17}$ GeV down to  $Q_0=10$ TeV.} 
\label{RGEfig}
%\end{minipage}
\end{center}
\end{figure}

We consider the two case of $Q_0=10$ TeV and $50$ TeV.
The parameter set of the first case (a) is given as
\begin{eqnarray}
\Lambda=10^{17} \ {\rm GeV} \ , \ Q_0=m_0=10 \ {\rm TeV} \ , \ m_{1/2}=6.2 \ {\rm TeV} \ , \
 \tan\beta=10 \ , \ A_0 =25.803  \ {\rm TeV}\ .
\end{eqnarray}
Here $m_{1/2}$ and $A_0$ are  tuned in order to obtain
 the proper $\lambda_H$ with the small $X_t(A_t)$,
 which gives $m_H=126$ GeV at the electroweak $m_H$ scale.
The parameter set of the second  case (b) is given as
\begin{eqnarray}
\Lambda=10^{16} \  {\rm GeV} \ , \ Q_0=m_0=50  \ {\rm TeV} \ , \ m_{1/2}=63.5 \ {\rm TeV} \ , \
 \tan\beta=4 \ , \ A_0 =109.993  \ {\rm TeV}\ .
 \end{eqnarray}
 These parameter sets are easily found  following from the numerical work in Ref.\cite{Delgado:2013gza}.
The obtained SUSY mass spectra at $Q_0$ are  summarized in Table 1,
where the top mass is sensitive to give the Higgs mass, and
we use $\overline m_t (m_t)=163.5\pm 2$ GeV \cite{UTfit,PDG}.
For the case (a), we show the running of  SUSY masses in the MSSM from $\Lambda$ down to  $Q_0$ in Figure \ref{RGEfig}
\cite{Martin:1997ns}.

As seen in Table 1, the first and second family squarks are degenerate in their masses, on the other hand,
the third ones  split due to the large RGE's effect.
Therefore, the mixing angle between the first and second family squarks vanishes, but the 
mixing angles between the first-third and the second-third family squarks are produced at the $Q_0$ scale.
The left-right mixing angle  between   $\tilde b_{L}$ and $\tilde b_{R}$ is given as 
\begin{eqnarray}
\theta\simeq \frac{m_b (A_b(Q_0)-\mu\tan\beta)}{m_{\tilde b_{L}}^2-m_{\tilde b_{R}}^2} \ .
\label{leftrightmixing}
\end{eqnarray}
It is noticed that the right-handed sbottom is heaver than the left-handed one.
The lightest squark is the right-handed stop.
Since we take the universal mass assumption for gauginos, $m_{1/2}$, the lightest gaugino is the Bino, $\tilde B$,
whose mass is $2.9$ TeV in the case of $Q_0=10$ TeV. That is the lightest supersymmetric particle (LSP) in our framework.
 Although these Wino and Bino mass values are  consistent 
with the recent experimental result of searching for EW-gaugino \cite{Aad:2014nua}, 
the Bino cannot be a candidate of the dark matter in this case  \cite{Harigaya:2014dwa,Harigaya:2014pqa}.  
In order to get the Wino dark matter, we should relax the universal mass assumption for gauginos.
However, this study does not affect our following  numerical results of the CP violation,
we do not discuss about the dark matter any more in this work.

%%%%%%%%%%%%%%%%%%%%%%%%%%%%%%%%%%%%%%%%%%%%%%%%%%%%%%%%%%%
%%%%%%%%%%%%%%%%%%%%%%%%%%%%%%%%%%%%%%%%%%%%%%%%%%%%%%%%%%%

%%%%%%%%%%%%%%%%%%%%%%%%%%%%%%%%%%%%%%%%%%%%%%%%%%%%%%%%%%
\begin{table}[h]
\begin{center}
\begin{tabular}{|c|l|l|}\hline
 &\quad Input at $\Lambda$ and $Q_0$ & \qquad  \qquad \qquad \quad Output at $Q_0$ \\
\hline
  & & \\
 Case  (a)&at $\Lambda=10^{17}$ GeV,  & $m_{\tilde g}=12.8$ TeV,  \ $m_{\tilde W}=5.2$ TeV,  
\ $m_{\tilde B}=2.9$ TeV\\ 
& \quad $m_0=10$ TeV,  & $m_{\tilde b_L}=m_{\tilde t_L}=12.2$ TeV  \\
& \quad $m_{1/2}=6.2$ TeV, &  $m_{\tilde b_R}=14.1$ TeV, \  $m_{\tilde t_R}=8.4$ TeV\\
& \quad  $A_0=25.803$ TeV; & $m_{\tilde s_L , \tilde d_L}=m_{\tilde c_L , \tilde u_L}=15.1$ TeV \\
& at $Q_0=10$ TeV, &   $m_{\tilde s_R,\tilde d_R}\simeq m_{\tilde c_R,\tilde u_R}=14.6$ TeV,  
\ $m_{{\cal H}}=13.7$ TeV  \\
&\quad  $\mu=10$ TeV,& $A_t=-1.2$ TeV, \   $A_b=5.1$ TeV, \  $X_t=-0.22$ \\
& \quad $\tan\beta=10$&   $\lambda_H=0.126$, \    $\theta=0.35^\circ$ \\
\hline
 & & \\
Case  (b)&at $\Lambda=10^{16}$ GeV,  & $m_{\tilde g}=115.6$ TeV,  \ $m_{\tilde W}=55.4$ TeV,  \ $m_{\tilde B}=33.45$ TeV\\ 
&\quad  $m_0=50$ TeV,  & $m_{\tilde b_L}=m_{\tilde t_L}=100.9$ TeV  \\
& \quad $m_{1/2}=63.5$ TeV, &  $m_{\tilde b_R}=104.0$ TeV, \  $m_{\tilde t_R}=83.2$ TeV\\
& \quad $A_0=109.993$ TeV; & $m_{\tilde s_L , \tilde d_L}=m_{\tilde c_L , \tilde u_L}=110.7$ TeV, \    $m_{\tilde s_R,\tilde d_R}=110.7$ TeV\\
&at $Q_0=50$ TeV,   &    $m_{\tilde c_R,\tilde u_R}=105.0$ TeV,  \ $m_{{\cal H}}=83.1$ TeV  \\
&\quad $\mu=50$ TeV,  & $A_t=-20.2$ TeV, \   $A_b=4.7$ TeV, \  $X_t=-0.65$ \\
&\quad  $\tan\beta=4$&   $\lambda_H=0.1007$, \    $\theta=0.05^\circ$ \\
\hline
\end{tabular}
\end{center}
\caption{Input parameters at $\Lambda$ and obtained the SUSY spectra in the cases of (a) and (b).}
\label{tab:inputparameters}
\end{table}
%%%%%%%%%%%%%%%%%%%%%%%%%%%%%%%%%%%%%%%%%%%%%%%%%%%%%%%%%%%%

%\begin{table}[t]
%\begin{center}
%\begin{tabular}{|l|}
%\hline
%$\alpha_s(M_Z)=0.1184$ \cite{PDG} \\
%$m_c(m_c)=1.275$ GeV \cite{PDG}\\
%$m_t(m_t)=163.5$ GeV $(\overline{{\text MS}})$ \cite{UTfit}\\
%$M_{B_s}=5.36677(24)$ GeV \cite{PDG}\\
%$\Delta M_{B_s}=(116.942 \pm 0.1564)\times 10^{-13}$ GeV \cite{Aaij:2013mpa}\\%[Moriond] \\
%$\Delta M_{B^0}=(3.337 \pm 0.033)\times 10^{-13}$ GeV \cite{PDG} \\
%$f_{B_s} = (233\pm 10)$ MeV \cite{UTfit}\\
%$f_{B_s}/f_{B^0} = 1.200\pm 0.02$ \cite{UTfit}\\
%$\xi_s=1.21(6)$ \cite{Buras:2008nn}\\
%$\lambda=0.2255(7)$ \cite{PDG}\\
%$|V_{cb}|=(4.12\pm0.11)\times 10^{-2}$  \cite{UTfit}\\
%$\eta_{cc}=1.43(23)$ \cite{Buras:2008nn}\\
%$\eta_{ct}=0.47(4)$ \cite{Buras:2008nn}\\
%$\eta_{tt}=0.5765(65)$ \cite{Buras:2008nn}\\
%$f_K = (156.1\pm1.1)$ MeV \cite{PDG}\\ %[Lattice HP] \\
%$\kappa_{\epsilon}=0.92(2)$ \cite{Buras:2008nn}\\
%\hline
%\end{tabular}
%\end{center}
%\caption{Input parameters in our calculation.}
%\label{tab:inputparameters}
%\end{table}

%%%%%%%%%%%%%%%%%%%%%%%%%%%%%
\section{Squark flavor mixing and CP violation}
\label{sec:CPVinSUSY}

\subsection{Squark flavor mixing}
\label{subsec:Squark flavor mixing}
Let us consider the $6\times 6$ squark mass matrix  $M_{\tilde q}$ in the super-CKM basis.
In order to move the mass eigenstate basis of squark masses,
we should diagonalize the mass matrix by rotation matrix $\Gamma _{G}^{(q)}$  as  
\begin{equation}
m_{\tilde q}^2=\Gamma _{G}^{(q)}M_{\tilde q}^2 \Gamma _{G}^{(q)  \dagger} \ ,
\end{equation}
where $\Gamma _{G}^{(q)}$ is the $6\times 6$ unitary matrix, and we decompose it into the $3\times 6$ matrices 
 as $\Gamma _{G}^{(q)}=(\Gamma _{GL}^{(q)}, \ \Gamma _{GR}^{(q))})^T$ in the following expressions:
\begin{align}
\Gamma _{GL}^{(d)}&=
\begin{pmatrix}
c_{13}^L & 0 & s_{13}^L e^{-i\phi_{13}^L} c_{\theta} & 0 & 0 & -s_{13}^L e^{-i\phi_{13}^L} s_{\theta} e^{i \phi} \\
-s_{23}^L s_{13}^L e^{i(\phi_{13}^L-\phi_{23}^L)} & c_{23}^L & s_{23}^Lc_{13}^Le^{-i\phi_{23}^L}c_{\theta} & 0 & 0 & -s_{23}^Lc_{13}^Le^{-i\phi_{23}^L}s_{\theta}e^{i\phi} \\
-s_{13}^Lc_{23}^Le^{i\phi_{13}^L} & -s_{23}^Le^{i\phi_{23}^L} &c_{13}^Lc_{23}^Lc_{\theta} & 0 & 0 & 
-c_{13}^Lc_{23}^Ls_\theta e^{i\phi}
\end{pmatrix}, \nonumber \\
\nonumber\\
\Gamma _{GR}^{(d)}&=
\begin{pmatrix}
0 & 0 & s_{13}^R s_\theta e^{-i \phi_{13}^R} e^{-i\phi } & c_{13}^R & 0 & s_{13}^R e^{-i \phi_{13}^R} c_{\theta} \\
0 & 0 & s_{23}^R c_{13}^R s_\theta e^{-i \phi_{23}^R} e^{-i\phi } & -s_{13}^R s_{23}^R e^{i(\phi_{13}^R-\phi_{23}^R)} & c_{23}^R & s_{23}^R c_{13}^R e^{-i \phi_{23}^R} c_{\theta} \\
0 & 0 & c_{13}^R c_{23}^R s_{\theta} e^{-i\phi } & -s_{13}^R c_{23}^R e^{i\phi_{13}^R} & -s_{23}^R e^{i \phi_{23}^R} & 
c_{13}^R c_{23}^R c_{\theta} \\
\end{pmatrix},
\label{mixing}
\end{align} 
where we use abbreviations $c_{ij}^{L,R}=\cos\theta_{ij}^{L,R}$, $s_{ij}^{L,R}=\sin\theta_{ij}^{L,R}$, 
 $c_\theta =\cos \theta$ and $s_\theta =\sin \theta$ in Eq. (\ref{leftrightmixing}).
Here $\theta$ is  the left-right mixing angle  between   $\tilde b_{L}$ and $\tilde b_{R}$.
 It is remarked that  we take $s_{12}^{L,R}=0$ due to the degenerate squark masses of the first and second families as discussed in the previous section. 

The gluino-squark-quark interaction is given as
\begin{equation}
\mathcal{L}_\text{int}(\tilde gq\tilde q)=-i\sqrt{2}g_s\sum _{\{ q\} }\widetilde q_i^*(T^a)
\overline{\widetilde{G}^a}\left [(\Gamma _{GL}^{(q)})_{ij}{P_L}
+(\Gamma _{GR}^{(q)})_{ij}{P_R}\right ]q_j+\text{h.c.}~,
\end{equation}
where $P_L=(1-\gamma _5)/2$, $P_R=(1+\gamma _5)/2$, and $\widetilde G^a$ denotes the gluino field,  
$q^i$ are three left-handed (i=1,2,3) and three right-handed quarks (i=4,5,6).
This interaction leads to the gluino-squark mediated flavor changing process 
with $\Delta F=2$ and  $\Delta F=1$ through the  box and  penguin diagrams.

\subsection{CP violation in $\Delta F=2$ and $\Delta F=1$ processes}

Taking account of the gluino-squark interaction, the dispersive part of meson mixing $M_{12}^{P}(P=K, B^0, B_s)$ are given as 
\begin{equation}
M_{12}^q=M_{12}^{q,\text{SM}}+M_{12}^{q,\text{SUSY}},
\end{equation}
where $M_{12}^{q,\text{SUSY}}$ are written by SUSY parameters in Eq.(\ref{mixing}) and its explicit formulation
is given in Appendix A.
The experimental data of $\Delta B=2$ process, the mass differences $\Delta M_{B^0}$ and $\Delta M_{Bs}$, 
and the CP-violating phases $\phi_d$ and $\phi_s$, give constraint  to the SUSY parameters in Eq.(\ref{mixing}).
We also consider the constraint from the CP-violating parameter in the $K$ meson, $\epsilon_K$,
and focus on the relation between $\epsilon_K$ and  $\sin (2 \beta)$, 
in which $\beta$ is one angle of  the unitarity triangle with respect to $B^0$.
%The SUSY contribution to the dispersive part of meson mixing $M_{12}^{P}(P=K, B^0, B_s)$ is parameterized as
%\begin{equation}
%M_{12}^q=M_{12}^{q,\text{SM}}+M_{12}^{q,\text{SUSY}}=
%M_{12}^{q,\text{SM}}(1+h_qe^{2i\sigma _q})~, %\quad (q=d,s)
%\end{equation}
%where $h_q$ and $\sigma _q$ denote magnitude and relative phase of new physics respectively,
%and they are written by SUSY parameters in Eq.\ref{mixing}.
%The explicit formulation of $M_{12}^{q,\text{SUSY}}$ is given in Appendix. 
%The experimental data of the mass difference of $K, B^0, B_s$ mesons give constraints 
%$\Delta M_P(P=K, B^0, B_s)$ are   

The indirect CP asymmetry in the semileptonic decays $B_q \to \mu^- X(q=d,s)$ leads to the nonzero
asymmetry $a^q_{sl}$ such as :
\begin{equation}
a^q_{sl}\equiv \frac{\Gamma (\bar B_q\rightarrow \mu ^+X)-\Gamma (B_q\rightarrow \mu ^-X)}
{\Gamma (\bar B_q\rightarrow \mu ^+X)+\Gamma (B_q\rightarrow \mu ^-X)}
\simeq \text{Im} \left (\frac{\Gamma _{12}^q}{M_{12}^q}\right )=
\frac{|\Gamma _{12}^q|}{|M_{12}^q|}\sin \phi^q_{sl}.
\end{equation}
The absorptive part of $B_q -{\bar B_q}$ system $\Gamma _{12}^q$ is dominated by the tree-level decay 
$b \to c{\bar c}s$ etc in the SM.
Therefore, we assume $\Gamma _{12}^q =\Gamma _{12}^{q, SM}$ in our calculation.
In the SM, the CP-violating phases are read \cite{Lenz:2011ti},
 \begin{equation}
  \phi^{s\rm SM}_{sl}=(3.84\pm 1.05)\times 10^{-3},  \qquad
  \phi^{d\rm SM}_{sl}=-(7.50\pm 2.44)\times 10^{-2},
 \end{equation}
which correspond to 
\begin{equation}
a^{s\rm SM}_{sl}=(1.9\pm 0.3)\times 10^{-5},  \qquad
a^{d\rm SM}_{sl}=-(4.1\pm 0.6)\times 10^{-4}.
  \label{semiCP}
 \end{equation}
 The recent experimental data of these CP asymmetries are given as 
\cite{Vesterinen:2013jia,PDG}
 \begin{equation}
a^{s}_{sl}=(-0.24\pm 0.54\pm 0.33)\times 10^{-2},  \qquad
a^{d}_{sl}=(-0.3\pm 2.1)\times 10^{-3}.
  \label{semiexp}
 \end{equation}

The time dependent CP asymmetries in non-leptonic decays are also interesting to search for the SUSY effect.
The $\Delta B=1$ transition amplitude is estimated by the effective Hamiltonian given as follows:
\begin{equation}
H_{eff}=\frac{4G_F}{\sqrt{2}}\left [\sum _{q'=u,c}V_{q'b}V_{q'q}^*
\sum _{i=1,2}C_iO_i^{(q')}-V_{tb}V_{tq}^*
\sum _{i=3-6,7\gamma ,8G}\left (C_iO_i+\widetilde C_i\widetilde O_i\right )\right ],
\label{hamiltonian}
\end{equation}
where $q=s,d$. The local operators are given as 
\begin{align}
&O_1^{(q')}=(\bar q_\alpha\gamma _\mu P_Lq_\beta')
(\bar q_\beta'\gamma ^\mu P_Lb_\alpha),
\qquad O_2^{(q')}=(\bar q_\alpha\gamma _\mu P_Lq_\alpha')
(\bar q_\beta'\gamma ^\mu P_Lb_\beta), \nonumber \\
&O_3=(\bar q_\alpha\gamma _\mu P_Lb_\alpha)\sum _Q(\bar Q_\beta\gamma ^\mu P_LQ_\beta),
\quad O_4=(\bar q_\alpha\gamma _\mu P_Lb_\beta)\sum _Q(\bar Q_\beta\gamma ^\mu P_LQ_\alpha), \nonumber \\
&O_5=(\bar q_\alpha\gamma _\mu P_Lb_\alpha)\sum _Q(\bar Q_\beta\gamma ^\mu P_RQ_\beta),
\quad O_6=(\bar q_\alpha\gamma _\mu P_Lb_\beta)\sum _Q(\bar Q_\beta\gamma ^\mu P_RQ_\alpha), \nonumber \\
&O_{7\gamma }=\frac{e}{16\pi ^2}m_b\bar q_\alpha\sigma ^{\mu \nu }P_Rb_\alpha
F_{\mu \nu }, 
\qquad O_{8G}=\frac{g_s}{16\pi ^2}m_b\bar q_\alpha\sigma ^{\mu \nu }
P_RT_{\alpha\beta}^ab_\beta G_{\mu \nu }^a,
\end{align}
where $\alpha $, $\beta $ are color indices, and $Q$ is taken to be $u,d,s,c$ quarks.
Here, the $C_i$ is the Wilson coefficient and includes SM contribution and gluino-squark one, such as 
$C_i =C_i^{\rm SM}+C_i^{\tilde g}$.  The $C_i^{\rm SM}$ is given in Ref.~\cite{Buchalla:1995vs}. 
The terms $\widetilde C_i$ and $\widetilde O_i$   are  obtained by replacing  L(R) with  R(L).
The magnetic penguin contribution $C_{7\gamma}$ and $C_{8 g}$ can be enhanced due to the left-right mixing.
For the $b\rightarrow s$ transition,
the gluino contributions to these the Wilson coefficients, $C_{7\gamma }$ and $C_{8G}$, are given as follows:
\begin{align}
C_{7\gamma }^{\tilde g}(m_{\tilde g}) &= 
\frac{8}{3}\frac{\sqrt{2}\alpha _s\pi }{2G_FV_{tb}V_{ts}^*} \nonumber \\
&\times \sum_{I=1}^6\Bigg [\frac{\big (\Gamma _{GL}^{(d)}\big )_{2I}^*}{m_{\tilde d_I}^2}
\left \{ \big (\Gamma _{GL}^{(d)}\big )_{3I}\left (-\frac{1}{3}F_2(x_{\tilde g}^I)\right )
+\frac{m_{\tilde g}}{m_b}\big (\Gamma _{GR}^{(d)}\big )_{3I}
\left (-\frac{1}{3}F_4(x_{\tilde g}^I)\right )\right \}\ ,
\end{align}
\begin{align}
C_{8G}^{\tilde g}(m_{\tilde g}) &= \frac{8}{3}\frac{\sqrt{2}\alpha _s\pi }{2G_FV_{tb}V_{ts}^*}
\Bigg [\sum_{I=1}^6 \frac{\big (\Gamma _{GL}^{(d)}\big )_{2I}^*}{m_{\tilde d_I}^2}
\left \{ \big (\Gamma _{GL}^{(d)}\big )_{3I}
\left (-\frac{9}{8}F_1(x_{\tilde g}^I)-\frac{1}{8}F_2(x_{\tilde g}^I)\right )\right .\nonumber \\
&\hspace{3.8cm}\left .+\frac{m_{\tilde g}}{m_b}\big (\Gamma _{GR}^{(d)}\big )_{3I}
\left (-\frac{9}{8}F_3(x_{\tilde g}^I)-\frac{1}{8}F_4(x_{\tilde g}^I)\right )\right \} \ , \
\end{align}
where  $F_{i}(x_{\tilde g}^{I})$ are the loop functions given in Appendix B
with $x_{\tilde g}^{I}=m_{\tilde g}^2 / m_{\tilde d_I}^2 (I=1-6)$.
We estimate $C_{7\gamma}^{\tilde g}$ and $C_{8G}^{\tilde g}$ at the $m_b$ scale including the effect of the leading order of QCD as follows~\cite{Buchalla:1995vs}:
\begin{equation}
\begin{split}
C_{7\gamma}^{\tilde g}(m_b)
&= \zeta C_{7\gamma}^{\tilde g}(m_{\tilde g})
+\frac{8}{3}(\eta-\zeta) C_{8G}^{\tilde g}(m_{\tilde g}), \cr
C_{8G}^{\tilde g}(m_b)
&=\eta C_{8G}^{\tilde g}(m_{\tilde g}),
\end{split}
\end{equation}
where 
\begin{eqnarray}
&\zeta=\left ( 
 \frac{\alpha_s(m_{\tilde b})}{\alpha_s(m_{\tilde g})} \right )^{\frac{16}{15}}
\left ( 
 \frac{\alpha_s(m_{\tilde g})}{\alpha_s(m_t)} \right )^{\frac{16}{21}}
 \left ( 
 \frac{\alpha_s(m_t)}{\alpha_s(m_b)} \right )^{\frac{16}{23}} \ , \nonumber\\
& \eta=\left ( 
 \frac{\alpha_s(m_{\tilde b})}{\alpha_s(m_{\tilde g})} \right )^{\frac{14}{15}}
\left ( 
 \frac{\alpha_s(m_{\tilde g})}{\alpha_s(m_t)} \right )^{\frac{14}{21}}
 \left ( 
 \frac{\alpha_s(m_t)}{\alpha_s(m_b)} \right )^{\frac{14}{23}} \ .
 \label{QCD}
 \end{eqnarray}
 In the expression of Eq.(\ref{QCD}),  the QCD correction is taken into account 
for the case of the gluino mass being  much smaller than the squark one \cite{Hisano:2013cqa}.
 
Now that we discuss the time dependent CP asymmetries of $B^0$ and $B_s$ decaying into the final state $f$,
which are defined as~\cite{Aushev:2010bq} :
\begin{equation}
S_f=\frac{2\text{Im}\lambda _{f}}{1+|\lambda_{f}|^2}\ ,
%C_f=\frac{1-|\lambda_{f}|^2}{1+|\lambda_{f}|^2}\ ,
\label{sf}
\end{equation}
where 
\begin{equation}
\lambda_{f}=\frac{q}{p} \frac{\bar A(\bar B_q^0\to f)}{A(B_q^0\to f)} , \qquad \qquad 
\frac{q}{p}\simeq \sqrt{\frac{M_{12}^{q*}}{M_{12}^{q}}} ,
\label{lambdaf}
\end{equation}
where $A(B_q^0\to f)$ is the decay amplitude in $B_q^0\to f$.
The time-dependent CP asymmetries $S_f$ are mixing induced CP asymmetry,
where $M_{12}^q$ and $A(B_q^0\to f)$ include the SUSY contributions in addition to the SM one. 

The time-dependent CP asymmetries in the $B^0\to J/\psi  K_S$ and $B_s\to J/\psi  \phi$ decays
are well known as the typical decay mode to  determine the unitarity triangle.
In this decays, we write $\lambda_{J/\psi  K_S}$ and $\lambda_{J/\psi  \phi}$
in terms of phase factors, respectively:
\begin{equation}
\lambda_{J/\psi  K_S}\equiv 
-e^{-i\phi _d}, \qquad \lambda _{J/\psi \phi } \equiv e^{-i\phi _s}.
\label{new}
\end{equation}
In the SM, the phase $\phi_d$ is given in terms of the  angle of  the unitarity triangle
 $\phi_1$ as $\phi_d=2\phi_1$.
On the other hand, $\phi_s$ is given as $\phi_s=-2\beta_s$,
in which $\beta_s$ is the one angle  of  the unitarity triangle in $B_s$.
Once $\phi_d$ is input, $\phi_s $  in the SM is predicted as \cite{Charles:2004jd}
\begin{equation}
\phi_s=-0.0363\pm 0.0017\ .
\end{equation}
%by putting  $|\bar \rho |=1$. 
If the SUSY contribution is non-negligible, $\phi_d=2\phi_1$ and $\phi_s=-2\beta_s$ are not satisfied any more.

The recent experimental data of these phases are \cite{Aaij:2013oba,Amhis:2012bh}
\begin{equation}
\sin \phi _d=0.679\pm 0.020\ , \qquad \phi_s=0.07\pm 0.09\pm 0.01 \ .
\label{phasedata}
\end{equation}
These experimental values also constrain the mixing angles and phases in Eq.(\ref{mixing}).

The $b \to s$ transition is  one-loop suppressed one in the SM, so the SUSY contribution to this process
is expected to be sizable.
In this point of view, we focus on the CP asymmetries in the $b \to s$ transition, 
$B^0 \to \phi K_S$ and  $B^0 \to \eta 'K^0$.
The CP asymmetries of $B^0 \to \phi K_S$ and $B^0 \to \eta 'K^0$ have been studied  for 
these twenty years~\cite{Khalil:2003bi,Endo:2004dc,Kagan:SLAC}.
In the SM, $S_{\phi K_S}$ and $S_{\eta 'K^0}$ are same to $S_{J/\psi K_S}$ within roughly
$10$\% accuracy because the CP phase comes from mixing $M_{12}^{d}$ in these mode.
Once taking account of the new physics contribution, 
the $S_{\phi K_S}$ and $S_{\eta 'K^0}$ are expected to be deviated from  $S_{ J/\psi K_S}$
because $B^0 \to  J/\psi K_S$ is the tree-level decay whereas $B^0 \to \phi K_S$ and $B^0 \to \eta 'K^0$
are one-loop suppressed one in the SM.
Recent experimental fit results of these CP asymmetries are reported by HFAG as follows~\cite{Amhis:2012bh}:  
\begin{equation}
S_{ J/\psi K_S}=0.679\pm 0.020 \ , \qquad 
S_{\phi K_S}= 0.74^{+0.11}_{-0.13}\ , \qquad 
S_{\eta 'K^0}= 0.59\pm 0.07\ .
%S_{K^{*}\gamma}= -0.16\pm{0.22}
\label{Sfdata}
\end{equation}
These values are may be regarded to be same within experimental error-bar and consistent with the SM prediction,
In other words, these experimental results give severe constraints to the squark flavor mixing angle between
the second-third families.

The CP asymmetries in $B^0 \to \phi K_S$ and $B^0 \to \eta 'K^0$ containing the SUSY contribution
are estimated in terms of $\lambda_f$ in Eq.({\ref{lambdaf}}):
\begin{align}
\lambda _{\phi K_S,~\eta 'K^0}&=-e^{-i\phi _d}\frac{\displaystyle \sum _{i=3-6,7\gamma ,8G}
\left (C_i \langle O_i \rangle +
\widetilde C_i \langle \widetilde O_i \rangle \right )}
{\displaystyle \sum _{i=3-6,7\gamma ,8G}\left (
C_i^{*}\langle O_i \rangle +\widetilde C_i^{*}\langle \widetilde O_i 
\rangle \right )}~,
\label{asymBd}
\end{align}
where $\langle O_i \rangle $ is the abbreviation for $\langle f|O_i|B^0\rangle $. 
It is known that $\langle \phi K_S|O_i|B^0\rangle =\langle \phi K_S|\widetilde O_i|B^0\rangle $ 
and $\langle \eta 'K^0|O_i|B^0\rangle =-\langle \eta 'K^0|\widetilde O_i|B^0\rangle $, 
because these final states have different parities~\cite{Khalil:2003bi,Endo:2004dc,Kagan:SLAC}. 
Then, the decay amplitudes of $f=\phi K_S$ and $f=\eta 'K^0$ are written
in terms of the dominant gluon penguin ones  $C_{8G}$ and $\tilde C_{8G}$ as follows: 
\begin{align}
\bar A(\bar B^0 \to \phi K_S)
& \propto C_{8G}(m_b) + {\tilde C}_{8G}(m_b), \nonumber \\
\bar A(\bar B^0 \to \eta '\bar K^0)
& \propto C_{8G}(m_b) - {\tilde C}_{8G}(m_b).
\end{align}
Since ${\tilde C}_{8G}(m_b)$ is suppressed compared to $C_{8G}(m_b)$ in the SM, 
the magnitudes of the time dependent CP asymmetries 
$S_f \ (f=J/\psi K_S, \ \phi K_S,\  \eta 'K^0)$ are almost same in the SM prediction. 
If the squark flavor mixing gives the unsuppressed ${\tilde C}_{8G}(m_b)$, 
 these CP asymmetries are expected to be deviated among them. 
%Therefore, those experimental data  give us the tight constraint for $C_{8G}(m_b)$ and ${\tilde C_{8G}}(m_b)$. 

In order to obtain precise results, we also take account of the small contributions 
from other Wilson coefficients $C_i~(i=3,4,5,6)$ and $\tilde C_i~(i=3,4,5,6)$ in our calculations. 
We estimate each hadronic matrix element by using the factorization relations in Ref.~\cite{Harnik:2002vs}: 
\begin{equation*}
\langle O_{3} \rangle =\langle O_{4} \rangle =\left( 1+\frac{1}{N_c} \right) \langle O_{5} \rangle,
\quad \langle O_{6} \rangle =\frac{1}{N_c}\langle O_{5} \rangle,
\end{equation*}
\begin{equation}
\langle O_{8G} \rangle =\frac{\alpha _s(m_b)}{8 \pi }
\left( -\frac{2 m_b}{ \sqrt{\langle q^2 \rangle }}\right )
\left( \langle O_4 \rangle +\langle O_6 \rangle -\frac{1}{N_c}(\langle O_3 \rangle 
+\langle O_5 \rangle )\right ), 
\end{equation}
where $\langle q^2 \rangle ={\rm 6.3~GeV^2}$ 
and $N_c=3$ is the number of colors. 
One may worry about the reliability of these naive factorization relations. 
However this approximation has been justified numerically 
in the relevant $b\to s$ transition as seen in the calculation of PQCD~\cite{Mishima:2003wm}. 

We also consider the SUSY contribution for the $b\to s\gamma$ decay. 
The $b\to s\gamma$ is sensitive to the magnetic penguin contribution $C_{7\gamma}$.
The branching ratio BR$(b\to s\gamma )$ is given as~\cite{Buras:1998raa} 
\begin{equation}
\frac{\text{BR}(b\to s\gamma )}
{\text{BR}(b\to ce\bar {\nu _e})}
=
\frac{|V_{ts}^*V_{tb}|^2}
{|V_{cb}|^2}
\frac{6 \alpha }{\pi f(z)}
(|C_{7\gamma }(m_b)|^2+|{\tilde C}_{7\gamma }(m_b)|^2),
\label{Brbqgamma}
\end{equation}
where 
\begin{equation}
f(z)=
1-8z+8z^3-z^4-12z^2 \text{ln}z~,\qquad 
z = \frac{m_{c,pole}^2}{m_{b,pole}^2}.
\end{equation}
%Here $C_{7\gamma }(m_b)$ and $\tilde{C}_{7\gamma }(m_b)$ include both contributions 
%from the SM and the gluino-squark mediated flavor changing process at the $m_b$ scale. 
The SM prediction including the next-to-next-to-leading order correction is given as \cite{Misiak:2006zs}
\begin{equation}
\text{BR}(b\to s\gamma )({\rm SM})=(3.15 \pm 0.23)\times 10^{-4},
\end{equation}
on the other hand, the experimental data are obtained as \cite{PDG} 
\begin{equation}
\text{BR}(b\to s\gamma )({\rm exp})=(3.53 \pm 0.24)\times 10^{-4}. 
\end{equation}
Therefore, we can examine the contribution of the gluino-squark mediated flavor-changing process to the $b\to s\gamma$ process.

In our analysis we also discuss the relation between $\epsilon_K$ and $ \sin 2\phi_1$, where $\phi_1$ is the
one angle of the unitarity triangle.
The parameter $\epsilon_K$ is given in the following theoretical formula
\begin{equation}
\epsilon_K
=
e^{i \phi_{\epsilon}} \sin{\phi_{\epsilon}} \left( \frac{\text{Im}(M_{12}^K)}{\Delta M_K}
+ \xi \right),  \qquad
\xi
=
\frac{\text{Im} A_0^K}{\text{Re} A_0^K} , \qquad
\phi_\epsilon=\tan^{-1}\left( \frac{2 \Delta M_K}{\Delta \Gamma_K} \right),
\end{equation}
with $A_0^K$ being the isospin zero amplitude in $K\to\pi\pi$ decays.
Here, $M_{12}^K$ is the dispersive part of the  $K^0-\bar{K^0}$ mixing, 
and $\Delta M_K$ is the mass difference in the neutral $K$ meson.
The effects of $\xi \ne 0$ and $\phi_{\epsilon} < \pi/4$ give suppression effect in $\epsilon_K$,
and it is parameterized as $\kappa_{\epsilon}$ and estimated by Buras and Guadagnoli \cite{Buras:2008nn} as:
\begin{equation}
\kappa_{\epsilon}
=
0.92 \pm 0.02 \ \ .
\end{equation}
%In the SM, the dispersive part $M_{12}^K$ is given as follows, 
%\begin{align}
%M_K^{12}
%&=
%\langle K| \mathcal{H}_{\Delta F=2} |\bar{K} \rangle \nonumber \\
%&=
%-\frac{4}{3}\left( \frac{G_F}{4 \pi} \right)^2 M_W^2 \hat{B}_K F_K^2 M_K \left( \eta_{cc} \lambda_c^2 E(x_c)
%+\eta_{tt} \lambda_t^2 E(x_t)
%+2 \eta_{ct} \lambda_c \lambda_t E(x_c,x_t) \right) ,
%\end{align}
%where $\lambda_c = V_{cs}V_{cd}^*,\  \lambda_t = V_{ts}V_{td}^*$,
%and $E(x)$'s are the one-loop functions \cite{Inami:1980fz}.
The $|\epsilon_K^{\text{SM}}|$ is given in terms of the Wolfenstein parameters $\lambda$, $\overline\rho$ and 
$\overline\eta$ as follows:
\begin{align}
|\epsilon_K^{\text{SM}}|
&=
\kappa_{\epsilon} C_{\epsilon} \hat{B}_K |V_{cb}|^2 \lambda^2 \bar{\eta} 
\left( |V_{cb}|^2 (1-\bar{\rho})\eta_{tt} E(x_t)
-\eta_{cc}E(x_c)
+\eta_{ct} E(x_c,x_t) \right) %\nonumber \\
%&=
%\kappa_{\epsilon} C_{\epsilon} \hat{B}_K |V_{cb}|^2 \lambda^2
%\left(
%\frac{1}{2} |V_{cb}|^2 R_t^2 \sin(2 \beta) \eta_{tt} E(x_t)
%+R_t \sin \beta (-\eta_{cc}E(x_c) +\eta_{ct} E(x_c,x_t))
%\right),
\label{epsilonKSM}
\end{align}
with
\begin{align}
 C_{\epsilon}=
\frac{G_F^2 F_K^2 m_K M_W^2}{6  \sqrt{2} \pi^2 \Delta M_K}.
\end{align}
It is easily found that 
$|\epsilon_K^{\text{SM}}|$ is proportional to $\sin(2\phi_1)$ because there is only one CP violating phase in the SM.
%and
%\begin{align}
%\bar\rho=\rho \  \left (1-\frac{1}{2}\lambda^2\right ),
%\qquad \bar\eta=\eta \  \left (1-\frac{1}{2}\lambda^2 \right ).
%\end{align}
%In Eq.(\ref{epsilonKSM}), we use the relation: 
%\begin{align}
%R_t \sin{\beta}
%= \bar{\eta}, \qquad
%R_t \cos{\beta}
%= 1- \bar{\rho},
%\end{align}
%where $R_t$ is
%\begin{align}
%R_t
%=
%\frac{1}{\lambda}
%\frac{|V_{td}|}{|V_{ts}|} 
%=
%\frac{1}{\lambda}
%\frac{F_{B_s}\sqrt{B_s}}{F_{B}\sqrt{B}}
%\sqrt{\frac{M_{{B_s}}}{M_{B^0}}}
%\sqrt{\frac{\Delta M_{B^0}^{\rm exp}}{\Delta M_{B_s}^{\rm exp}}}.
%\end{align}
%As seen in Eq.(\ref{epsilonKSM}),  $|\epsilon_K^{\text{SM}}|$ is given 
%in terms of  $\sin(2\beta)$  because
% there is only one CP violating phase in the SM.
Therefore, the observed value of $S_{J /\psi K_S}$, which correspond to $\sin(2\phi_1)$,
should be correlated with $|\epsilon_K|$ in the SM.
%In other words, one should test numerically  the overlapping among $|\epsilon_K|$, ${\Delta M_{B^0}}/{\Delta M_{B_s}}$
%and $\sin(2\beta)$ in the unitarity triangle.
According to the recent experimental results, 
it is found that the consistency between the SM prediction and the experimental data in $\sin(2\phi_1)$ and  
$|\epsilon_K^{\text{SM}}/\hat B_K|$ is marginal.
This fact was pointed out by Buras and Guadagnoli \cite{Buras:2008nn} 
and called as the tension between $|\epsilon_K|$ and $\sin(2\phi_1)$.
Note that $|\epsilon_K^{\text{SM}}|$ also depends on the non-perturbative parameter $\hat{B}_K$ in Eq.(\ref{epsilonKSM}). 
Recently, the error of this parameter shrank dramatically in the lattice calculations \cite{Bae:2013lja}.
In our calculation we use the  updated value by the Flavor Lattice Averaging Group \cite{Aoki:2013ldr}:
\begin{equation}
\hat B_K= 0.766 \pm 0.010\ \ .
\label{BK}
\end{equation}
We can calculate $|\epsilon_K^{\text{SM}}|$ for the fixed  $\sin(2\phi_1)$ by inputting this value.

Considering the effect of the squark flavor mixing in both $|\epsilon_K|$ and $S_{J /\psi K_S}$,  this tension can be relaxed though the gluino-squark interaction.   
Then,  $\epsilon_K$ is expressed as:
 \begin{align}
  \epsilon_K=\epsilon_K^{\text{SM}}+\epsilon_K^{\rm SUSY},
  \label{epsilon}
  \end{align}
where $\epsilon_K^{\rm  SUSY}$ is induced by the imaginary part of the gluino-squark box diagram, 
which is presented in Appendix A. 
Since $s_{12}^{L(R)}$ vanishes in our scheme, 
 $\epsilon_K^{\tilde g}$ is given in the second order of the squark mixing
  $s_{13}^{L(R)}\times s_{23}^{L(R)}$. 
%because the first and second families are decoupled in the  gluino-squark box diagrams.
%We should also modify $R_t$ as follows:
% \begin{align}
%R_t
%=
%\frac{1}{\lambda}
%\frac{F_{B_s}\sqrt{B_s}}{F_{B}\sqrt{B}}
%\sqrt{\frac{M_{B_s}}{M_B}}
%\sqrt{\frac{\Delta M_{B^0}^{\rm exp}}{\Delta M_{B_s}^{\rm exp}}}
%\sqrt{\frac{C_s}{C_d}},
%\end{align}
%where
%\begin{align}
%C_q= 1+h_q e^{2i\sigma _q}, \quad (q=d, \ s).
%\end{align}
%%%%%%%%%%%%%%%%%%%%%%%%%%%
%where, $\xi_s\equiv\frac{F_{B_s}\sqrt{\hat{B}_s}}{F_{B_d}\sqrt{\hat{B}_d}}$, 
%$\langle B_q|H^{\text{full}}_{\Delta F=2}|\bar{B}_q\rangle\equiv A^{\text{full}}_q e^{2 i \beta^{\text{full}}_q}$,
%$A^{\text{full}}_q = A^{\text{SM}}_q C_q$ .
%%%%%%%%%%%%%%%%%%%%%%%%%%%%
%%%%%%%%%%%%%%%%%%%%%%%%%%%%%%%%%%%%%%%%%%%%%%%%%%%%5

%%%%%%%%%%%%%%%%%%%%%%%%%%%%%%%%%%%

In addition to the above CP violating processes, the neutron EDM is also sensitive to
the CP-violating phase of the squark mixing through cEDM of the strange quark. 
The experimental upper bound of the  electric dipole moment of the neutron
provides us the upper-bound of cEDM of the strange quark 
\cite{Hisano:2003iw}-\cite{Fuyuto:2012yf}. 
The cEDM of the strange quark $d_{s}^C$ comes from the gluino-squark interactions is given in Appendix C.
The bound on the cEDM of the strange quark 
is estimated as \cite{Fuyuto:2012yf} from the experimental upper bound of the neutron EDM as follows:
\begin{equation}
 e|d_s^C|<0.5\times 10^{-25} \ \text{ecm}.
 \label{cedm}
 \end{equation}
 %\cite{Hisano:2003iw,Hisano:2004tf,Hisano:2008hn,Fuyuto:2012yf}. 
This bound also give severe constraints for phases of the mixing parameters of Eq.(\ref{mixing}).

%%%%%%%%%%%%%%%%%%%%%%%%%%%%%%%%%%%%%%%%%%%%%%%%%%%%%%%%%%%%%%%%%%%%%%%%%%%%%%%
%%%%%%%%%%%%%%%%%%%%%%%%%%%%%%%%%%%%%%%%%%%%%%%%%%%%%%%%%%%%%%%%%%%%%%%%%%%%%%%

\section{Numerical  results}
In this section we show our numerical results.
At the first step, we constrain the squark flavor mixing parameters in Eq. (\ref{mixing}) from the experimental data of the CP violation $\epsilon_K$, $\phi_d$ and $\phi_s$, and the mass difference $\Delta M_{B^0}$ and $\Delta M_{B_s}$ comprehensively.
We have nine free  parameters, in which 
there are four mixing angles $\theta_{13}^{L(R)}$ and  $\theta_{23}^{L(R)}$, 
 five phase $\phi_{13}^{L(R)}$, $\phi_{23}^{L(R)}$,  $\phi$.
 In our analyses, we reduce the number of parameters
by taking  $\theta_{ij}^{\rm L} =\theta_{ij}^{\rm R}$ for simplicity,
but we also discuss the case where this assumption is broken in the estimate of
 $\epsilon_K$ and  the cEDM of the strange quark.
Moreover,  Wolfenstein parameters $\bar\rho$, $\bar\eta$
are free ones, which are determined by our numerical analyses.
Other relevant input parameters such as quark masses $m_c$, $m_b$,  the CKM matrix elements
 $V_{us} $, $V_{cb}$  and  $f_B$, $f_K$,  etc. are shown in our previous paper
Ref. \cite{Shimizu:2013jia}, which are referred from the PDG \cite{PDG} and the UTfit Collaboration \cite{UTfit}.

The uncertainties of these input parameters determine the predicted range of the SUSY contribution
for the CP violations, $\Delta M_{B^0}$ and $\Delta M_{B_s}$.
For example, the predicted range of the SUSY contribution for $\epsilon_K$ 
mainly  comes from the uncertainties of $\hat B_K$, $|V_{cb}|$ and  $m_t$ in addition to
the observed error bar of   $|\epsilon_K|$.  
If these uncertainties will be reduced in the future, the predicted range of the CP violation is improved.

At the second step, we predict the deviations of the time dependent CP asymmetries  $S_f$ and
the semileptonic CP asymmetries $a^q_{sl}(q=d,s)$  from the SM taking account of the contribution of the gluino-squark interaction.
The SUSY effect on the cEDM of the strange quark is also discussed.

In our analysis,  we scan the mixing angles  $s_{ij}^{\rm L(R)}$ and  phases in Eq. (\ref{mixing})
in the region of $0 \sim 0.5$ and $0 \sim 2\pi$,  respectively.
At first, we show the analysis  in the case of the SUSY scale $Q_0= 10 \ {\rm TeV}$ in detail,
and then,  we also discuss the numerical results in the the case $Q_0= 50 \ {\rm TeV}$.

%%%%%%%%%%%%%%%%%%%%%%%%%%%%%%%%%%%%%%%%%%%%%%%%%%%%%%%%%%%
\begin{figure}[t]
\begin{minipage}[]{0.45\linewidth}
\includegraphics[width=7.5cm]{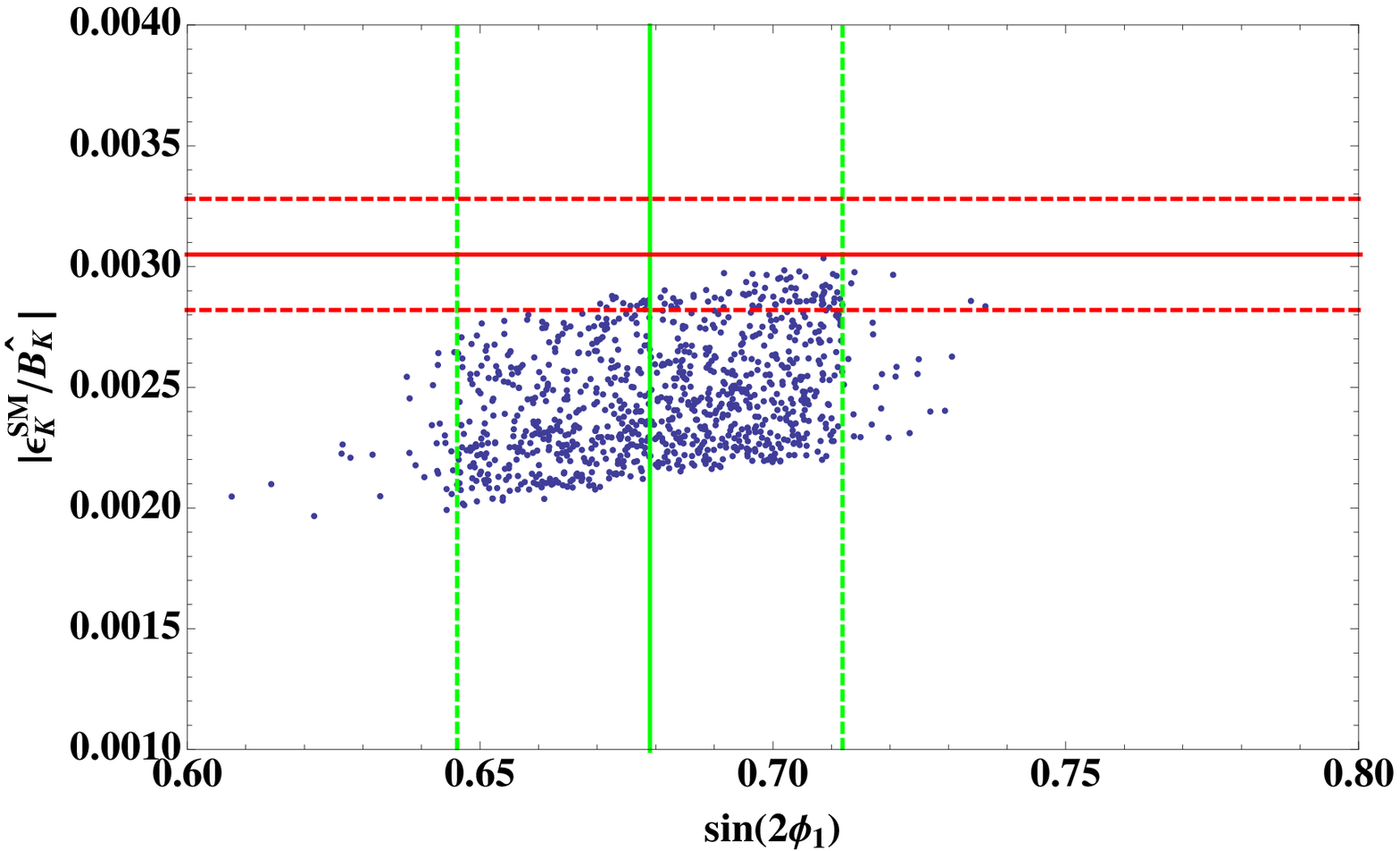}
\hspace{1cm}
\caption{Predicted region on $\sin(2\phi_1)-|\epsilon_K^{\text{SM}}|/\hat B_K$ plane for $Q_0=10$ TeV. 
Vertical and horizontal dashed lines denote the experimental allowed region with $90\%$C.L. Vertical and horizontal solid lines denote observed  central values. } 
\label{tension}
\end{minipage}
\hspace{1cm}
\begin{minipage}[]{0.45\linewidth}
\vspace{-1.5cm}
\includegraphics[width=7.5cm]{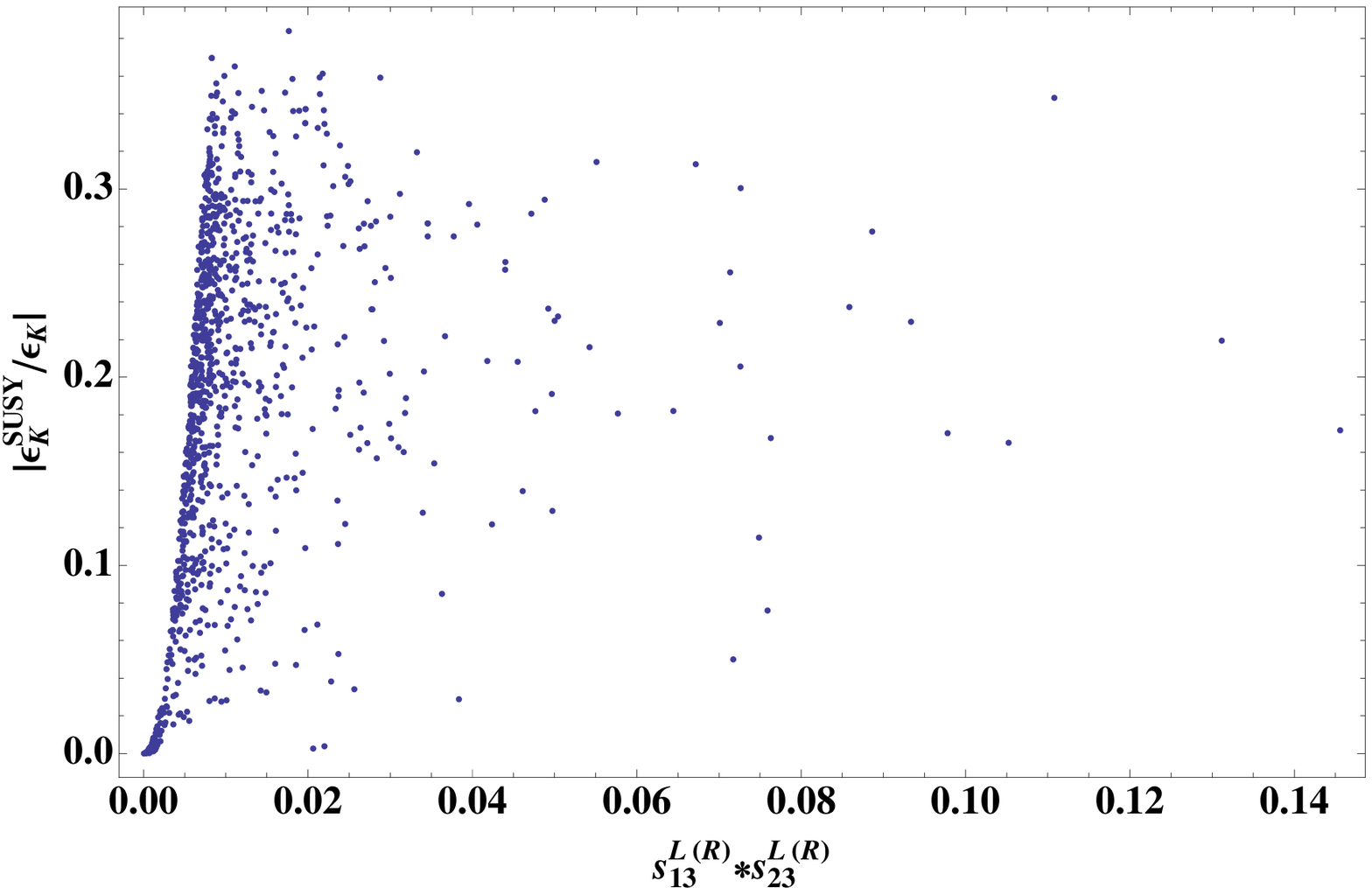}
\hspace{1cm}
\caption{The  predicted $|\epsilon_K^{\rm SUSY} / \epsilon_K|$ versus
$s_{13}^{\rm L(R)} \times s_{23}^{\rm L(R)}$ for $Q_0=10$ TeV. } 
\label{SUSYepsilon}
\end{minipage}
\end{figure}
%%%%%%%%%%%%%%%%%%%%%%%%%%%%%%%%%%%%%%%%%%%%%%%%%%%%%%%%%%
%%%%%%%%%%%%%%%%%%%%%%%%%%%%%%%%%%%%%%%%%%%%%%%%%%%%%%%%%%
\begin{figure}[!h]
\begin{minipage}[]{0.45\linewidth}
\includegraphics[width=7.8cm]{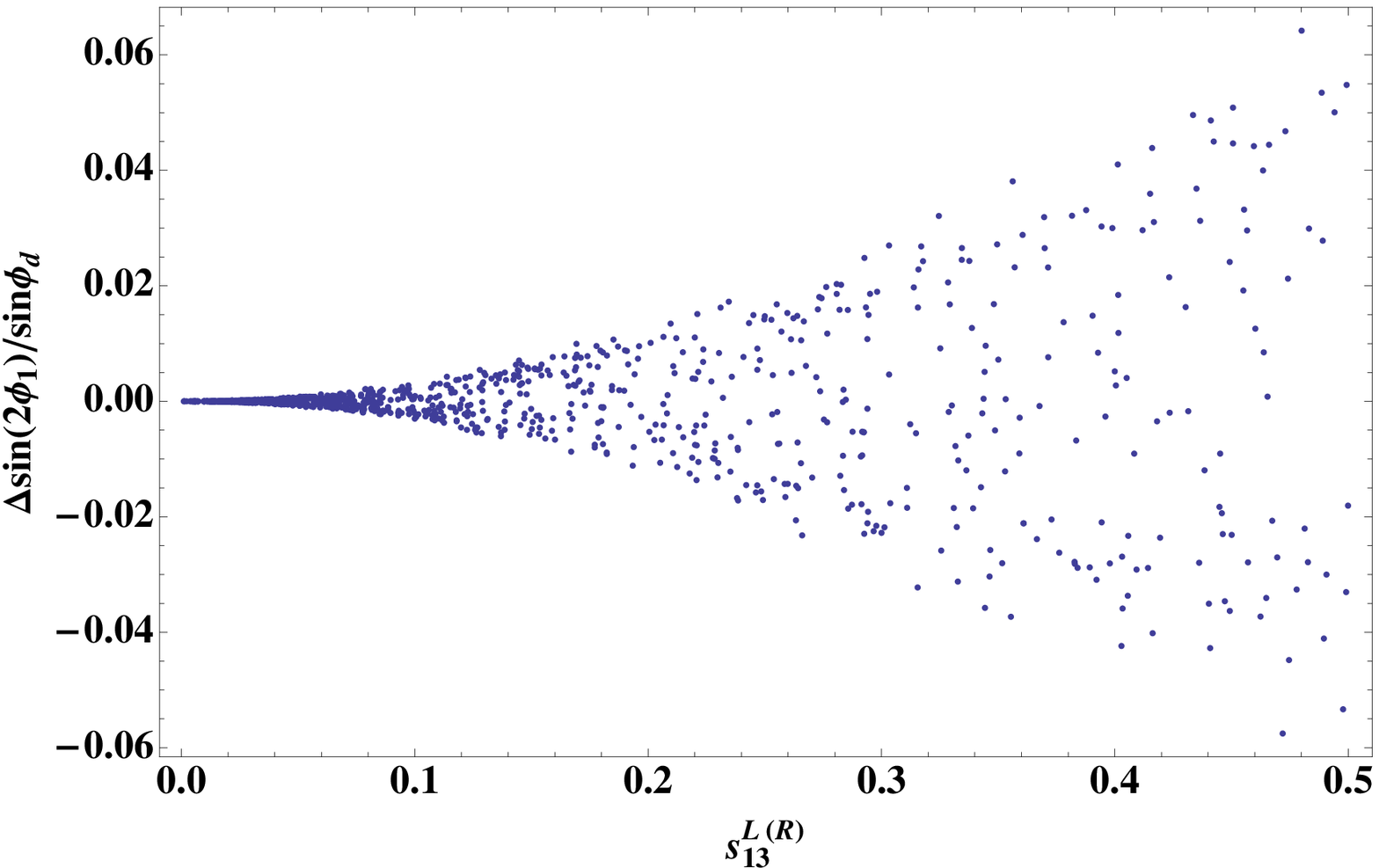}
\hspace{1cm}
\caption{The deviation of $\sin\phi_d$ from $\sin 2\phi_1$ versus $s_{13}^{L(R)}$.} 
\label{DeltaSind}
\end{minipage}
\hspace{1cm}
\begin{minipage}[]{0.45\linewidth}
\includegraphics[width=7.5cm]{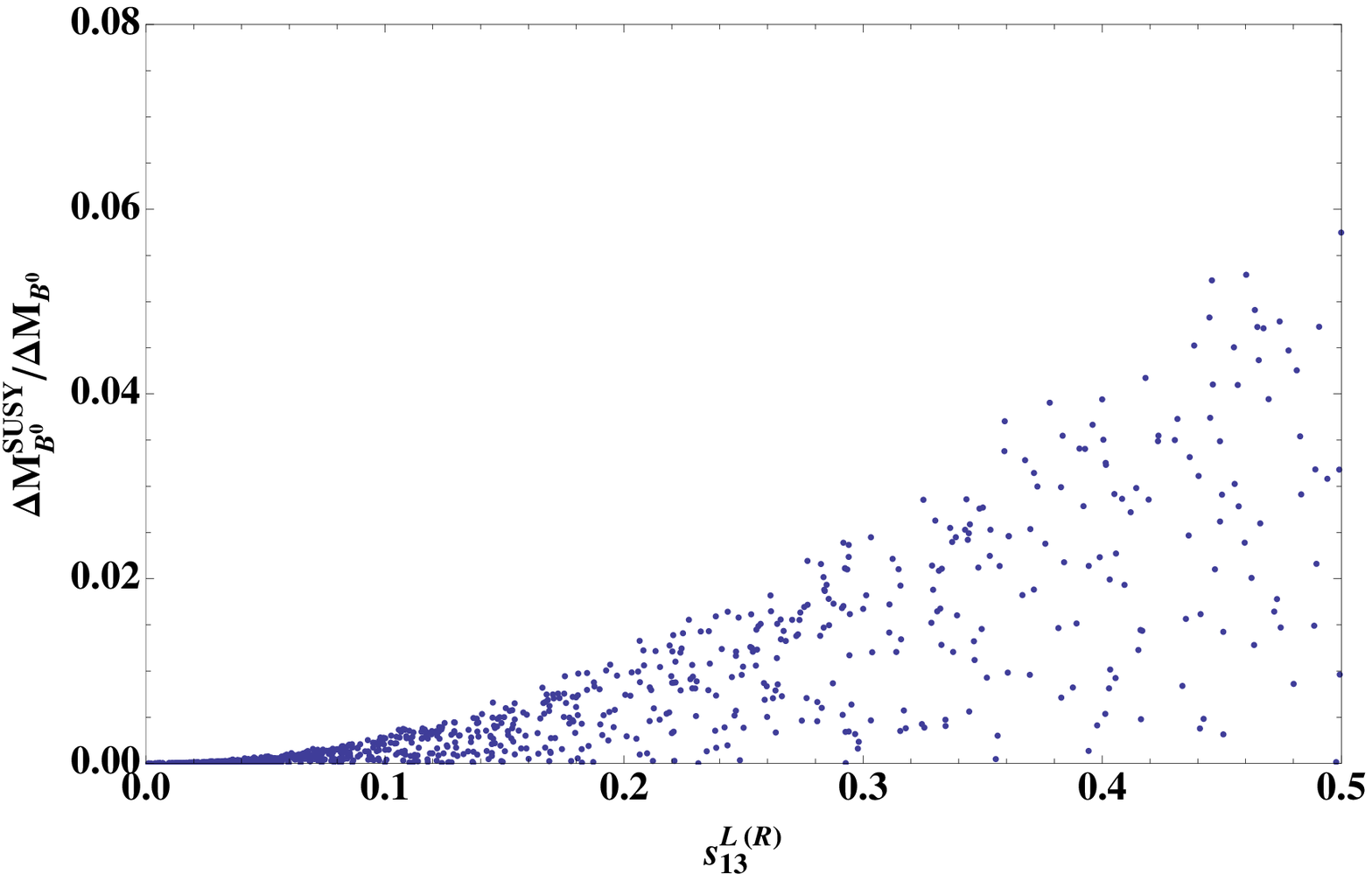}
\hspace{1cm}
\caption{The SUSY contribution to   $\Delta M_{B^0}$  versus $s_{13}^{L(R)}$.} 
\label{DeltaMd}
\end{minipage}
\end{figure}
%%%%%%%%%%%%%%%%%%%%%%%%%%%%%%%%%%%%%%%%%%%%%%%%%%%%%%%%%%%

Let us start with discussing  the gluino-squark interaction effect on the $\Delta F=2$ processes,
$\epsilon_K$, $\Delta M_{B^0}$ and $\Delta M_{B_s}$, where
the squark and gluino mass spectrum in Table 1 is input.
We show the allowed region on the plane of 
  $\sin (2\phi_1)$ and  $|\epsilon_K^{\rm SM}/{\hat B_K}|$ in Fig. \ref{tension}.
  When we add the contribution of the gluino-squark interaction,  $\epsilon_K^{\rm SUSY}$,
  the allowed region of  $\sin (2\phi_1)$ and  $|\epsilon_K^{\rm SM}/{\hat B_K}|$ converge within the experimental
  error-bar, where  $\phi_d$ is not $2\phi_1$ any more as discussed below Eq.(\ref{new}).
The Figure \ref{SUSYepsilon}  shows the $s_{13}^{\rm L(R)} \times s_{23}^{\rm L(R)}$ dependence of the SUSY contribution 
for  $\epsilon_K$, that is  $|\epsilon_K^{\rm SUSY} / \epsilon_K|$.
It is found that the SUSY contribution could be large up to $40\%$.
It is remarked that   $\epsilon_K$ is sensitive to the gluino-squark interaction even if the SUSY scale is $10$ TeV.

We show the SUSY contribution to the CP violating phase $\phi_d$ versus $s_{13}^{L(R)}$ in Figure \ref{DeltaSind}, 
where we define $\Delta \sin 2\phi_1\equiv \sin \phi_d-\sin 2\phi_1$, which vanishes in the SM.
The  $\sin\phi_d$ could be deviated from the SM  in $6\%$ as seen in this figure. 
We present  the SUSY contribution to the mass difference  $\Delta M_{B^0}$   versus $s_{13}^{L(R)}$
in Figure \ref{DeltaMd}.
It is remarked that the SUSY contribution could be also $6\%$ in the $\Delta M_{B^0}$.

%%%%%%%%%%%%%%%%%%%%%%%%%%%%%%%%%%%%%%%%%%%%%%%%%%%%%%%%%%% 

%%%%%%%%%%%%%%%%%%%%%%%%%%%%%%%%%%%%%%%%%%%%%%%%%%%%%%%%%%
\begin{figure}[t]
\begin{minipage}[]{0.45\linewidth}
\includegraphics[width=7.8cm]{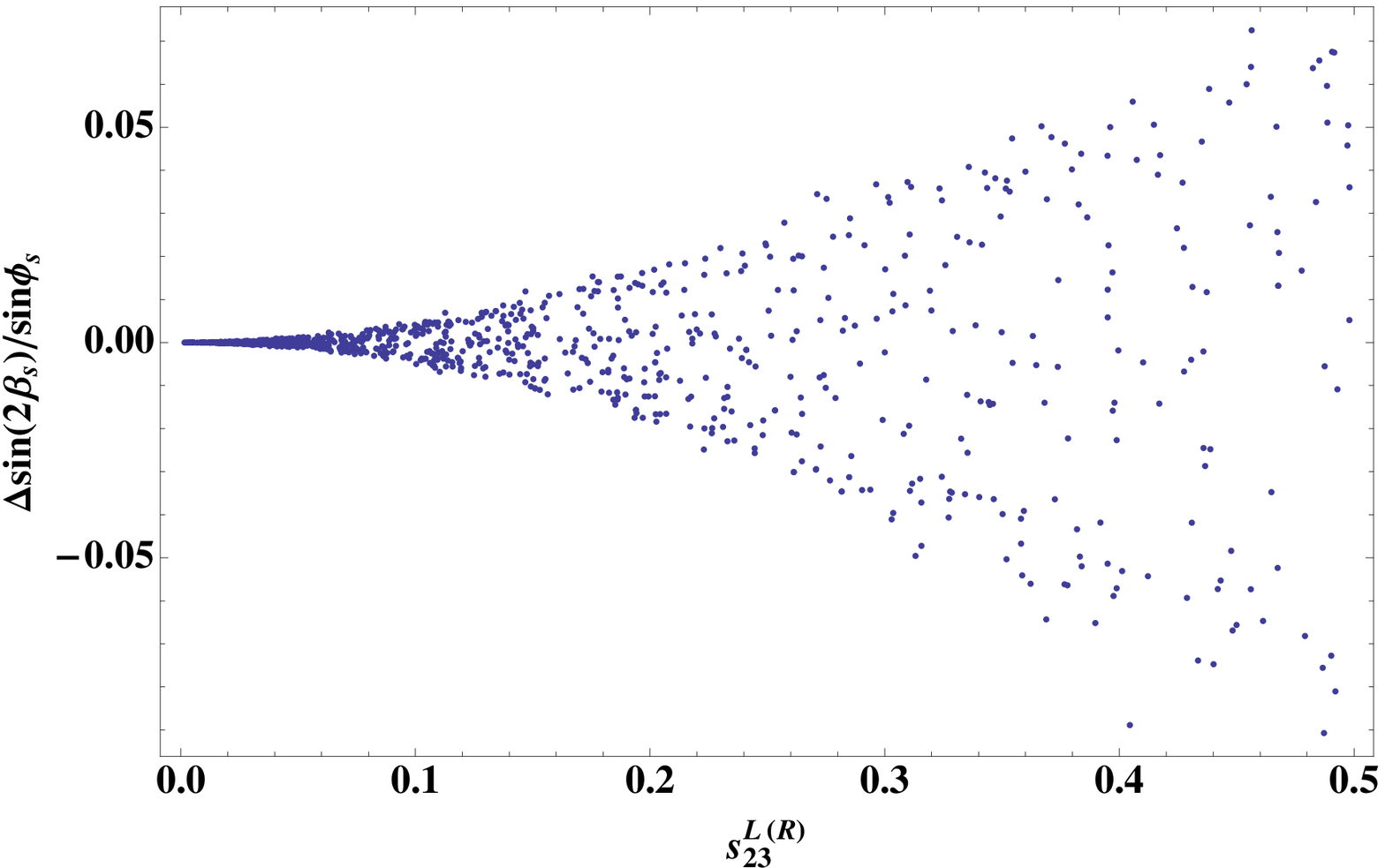}
\hspace{1cm}
\caption{The deviation of $\sin\phi_s$ from $\sin 2\beta_s$ versus $s_{23}^{L(R)}$.} 
\label{DeltaSins}
\end{minipage}
\hspace{1cm}
\begin{minipage}[]{0.45\linewidth}
\includegraphics[width=7.5cm]{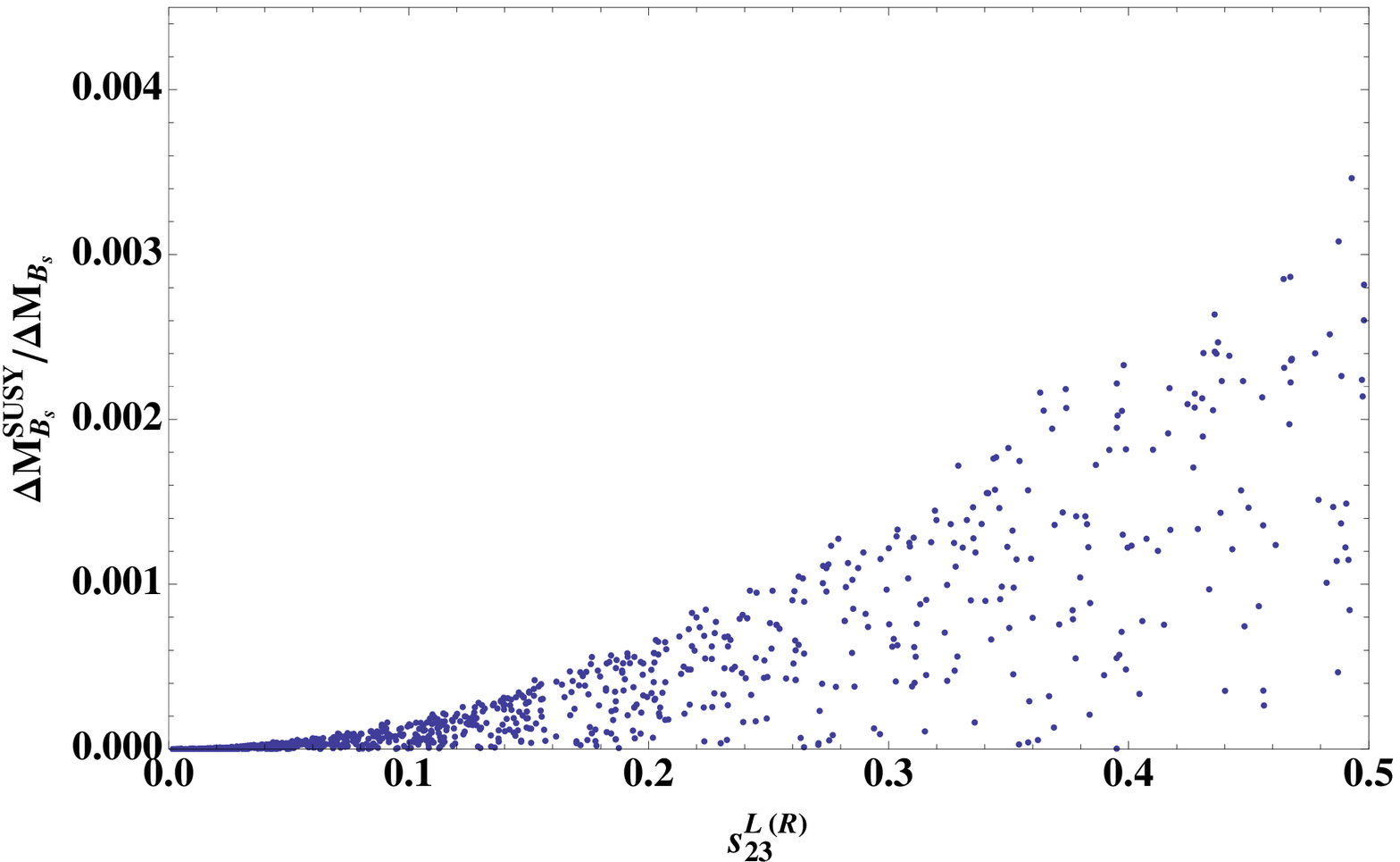}
\hspace{1cm}
\caption{The SUSY contribution to   $\Delta M_{B_s}$  versus $s_{23}^{L(R)}$.} 
\label{DeltaMs}
\end{minipage}
\end{figure}
%%%%%%%%%%%%%%%%%%%%%%%%%%%%%%%%%%%%%%%%%%%%%%%%%%%%%%%%%%%
%%%%%%%%%%%%%%%%%%%%%%%%%%%%%%%%%%%%%%%%%%%%%%%%%%%%%%%%%%%
\begin{figure}[!h]
\begin{minipage}[]{0.45\linewidth}
\includegraphics[width=7.8cm]{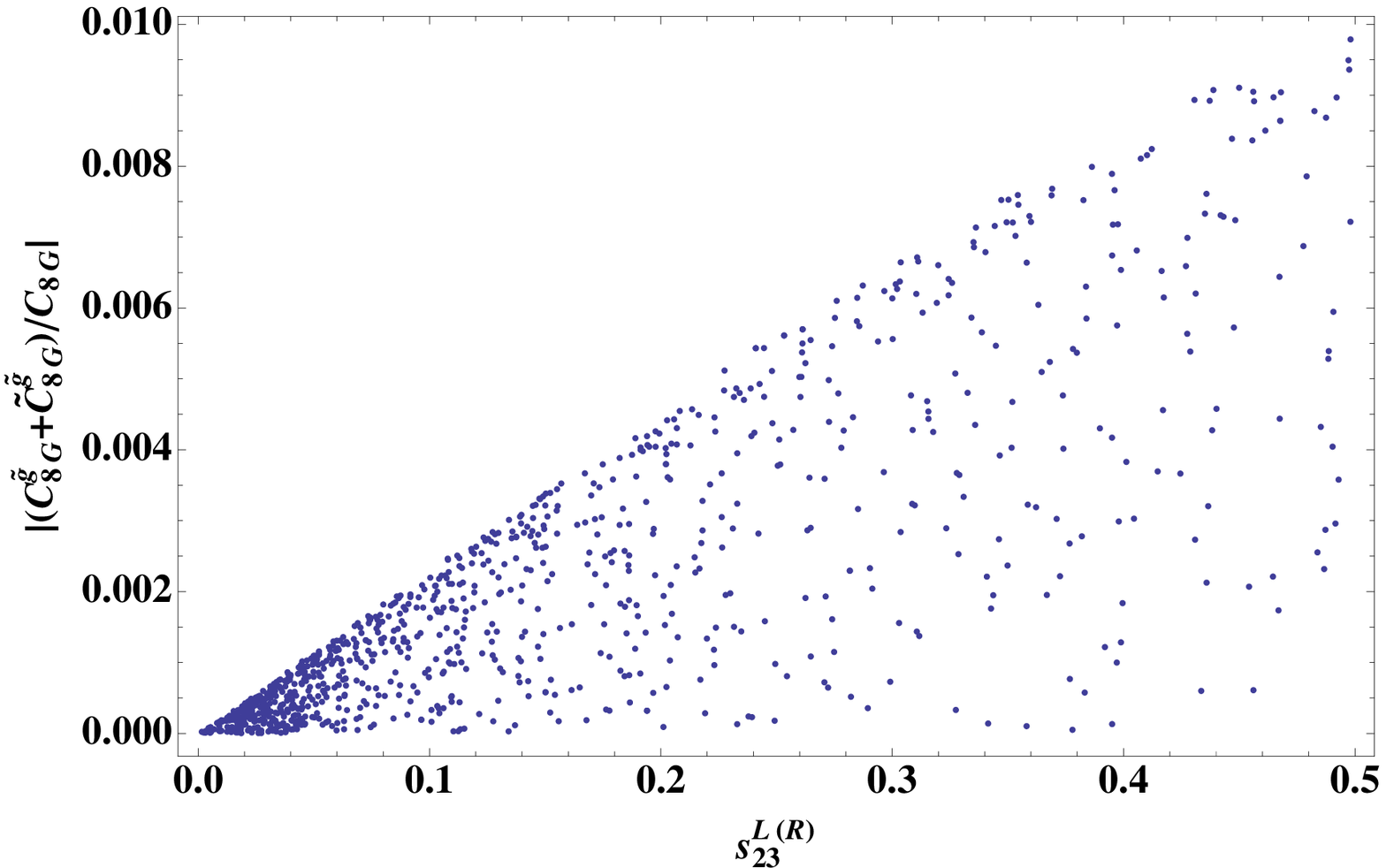}
\hspace{1cm}
\caption{ The predicted $| (C_{8G}^{\tilde g}+\tilde C_{8G}^{\tilde g})/C_{8G}|$ versus $s_{23}^{L(R)}$.} 
\label{C8}
\end{minipage}
\hspace{1cm}
\begin{minipage}[]{0.45\linewidth}
\includegraphics[width=7.5cm]{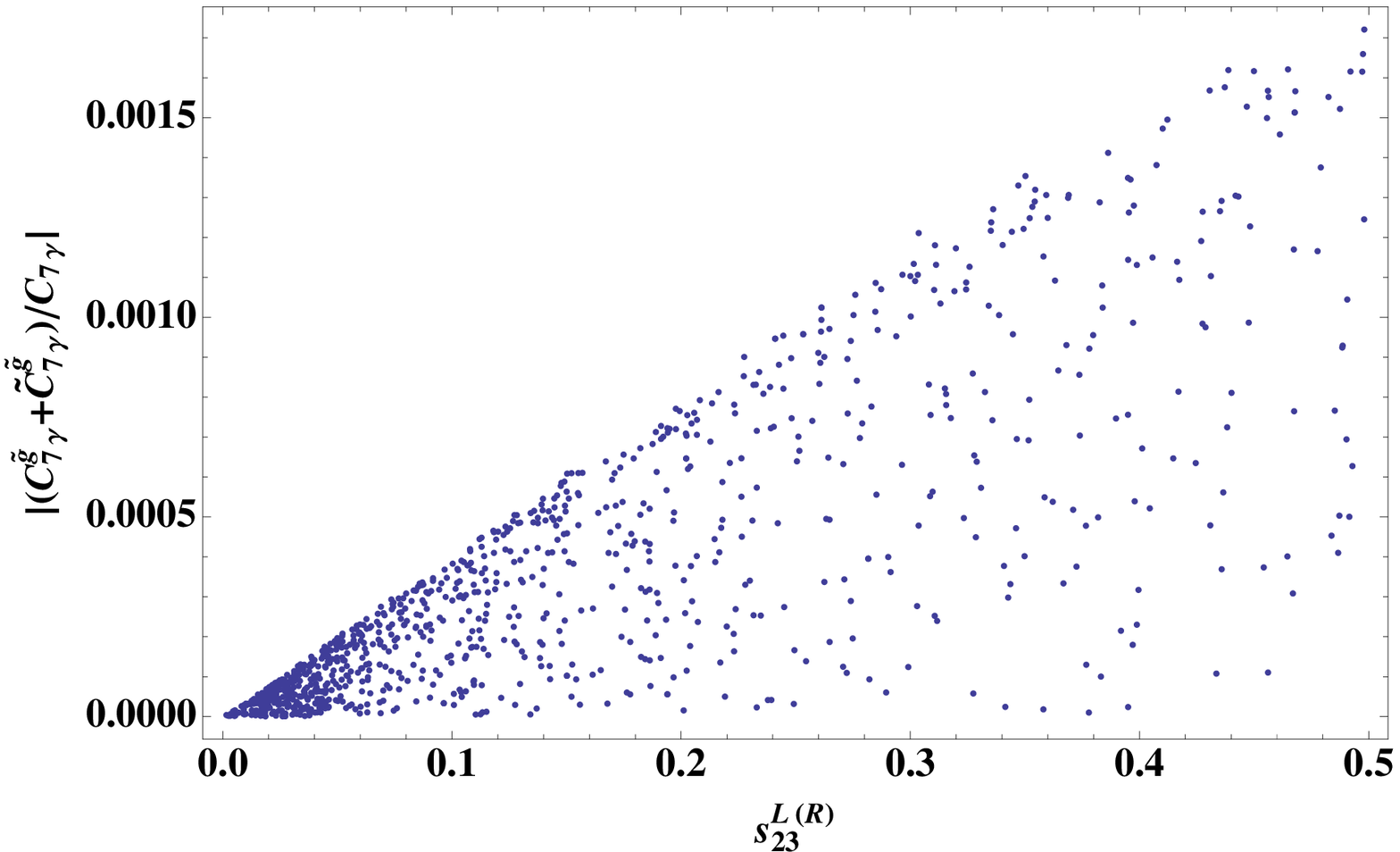}
\hspace{1cm}
\caption{The predicted $|(\tilde C_{7\gamma}^{\tilde g}+C_{7\gamma}^{\tilde g})/C_{7\gamma}|$  versus $s_{23}^{L(R)}$.} 
\label{C7}
\end{minipage}
\end{figure}
%%%%%%%%%%%%%%%%%%%%%%%%%%%%%%%%%%%%%%%%%%%%%%%%%%%%%%%%%%%
%%%%%%%%%%%%%%%%%%%%%%%%%%%%%%%%%%%%%%%%%%%%%%%%%%%%%%%%%%%
%\begin{figure}[h]
%\begin{minipage}[]{0.45\linewidth}
%\includegraphics[width=7.8cm]{Sphiks-Setak.eps}
%\hspace{1cm}
%\caption{$90\%$C.L. } 
%\label{Sphi}
%\end{minipage}
%\hspace{1cm}
%\begin{minipage}[]{0.45\linewidth}
%\includegraphics[width=7.5cm]{fig11.eps}
%\hspace{1cm}
%\caption{one sigma} 
%\label{Sphiratio}
%\end{minipage}
%\end{figure}
%%%%%%%%%%%%%%%%%%%%%%%%%%%%%%%%%%%%%%%%%%%%%%%%%%%%%%%%%%%
We show the SUSY contribution to the CP violating phase $\phi_s$ versus $s_{23}^{L(R)}$ in Figure \ref{DeltaSins}, 
where we define $\Delta \sin 2\beta_s\equiv \sin \phi_s-\sin 2\beta_s$, which vanishes in the SM.
It is found that the deviation of $\sin\phi_s$ from $\sin 2\beta_s$ is at most $8\%$.
As seen in Figure \ref{DeltaMs}, the SUSY contribution  for $\Delta M_{B_s}$ is very small,  ${\cal O}(0.4)\%$.

Let us discuss  the $b\rightarrow s$ transitions.
Under the constraints of the experimental data  $\epsilon_K$, $\phi_d$ and $\phi_s$,  $\Delta M_{B^0}$ and $\Delta M_{B_s}$,
 we can predict the magnitude of the Wilson coefficients $C^{\tilde g}_i$ and  $\tilde C^{\tilde g}_i$, which 
 give us the deviation from the SM predicted values.
 We show the ratio $| (C_{8G}^{\tilde g}+\tilde C_{8G}^{\tilde g})/C_{8G}|$ versus $s_{23}^{L(R)}$ 
in Figure \ref{C8}.
Thus  $C_{8G}^{\tilde g}$ is at most  $1\%$ because of the  small left-right mixing $\theta=0.35^\circ$ as seen in Table 1.
We also show the predicted  $|(\tilde C_{7\gamma}^{\tilde g}+C_{7\gamma}^{\tilde g})/C_{7\gamma}|$ 
  in Figure \ref{C7}. This magnitude is much smaller than  the case of $C_{8G}^{\tilde g}$, about $0.15\%$. 
Thus  $C_{7\gamma}^{\tilde g}$ do not affect  the 
branching ratio of the $b\rightarrow s\gamma$ decay in Eq.(\ref{Brbqgamma}). 
 %%%%%%%%%%%%%%%%%%%%%%%%%%%%%%%%%%%%%%%%%%%%%%%%%%%%%%%%%%
\begin{figure}[t]
\begin{minipage}[]{0.45\linewidth}
\includegraphics[width=7.8cm]{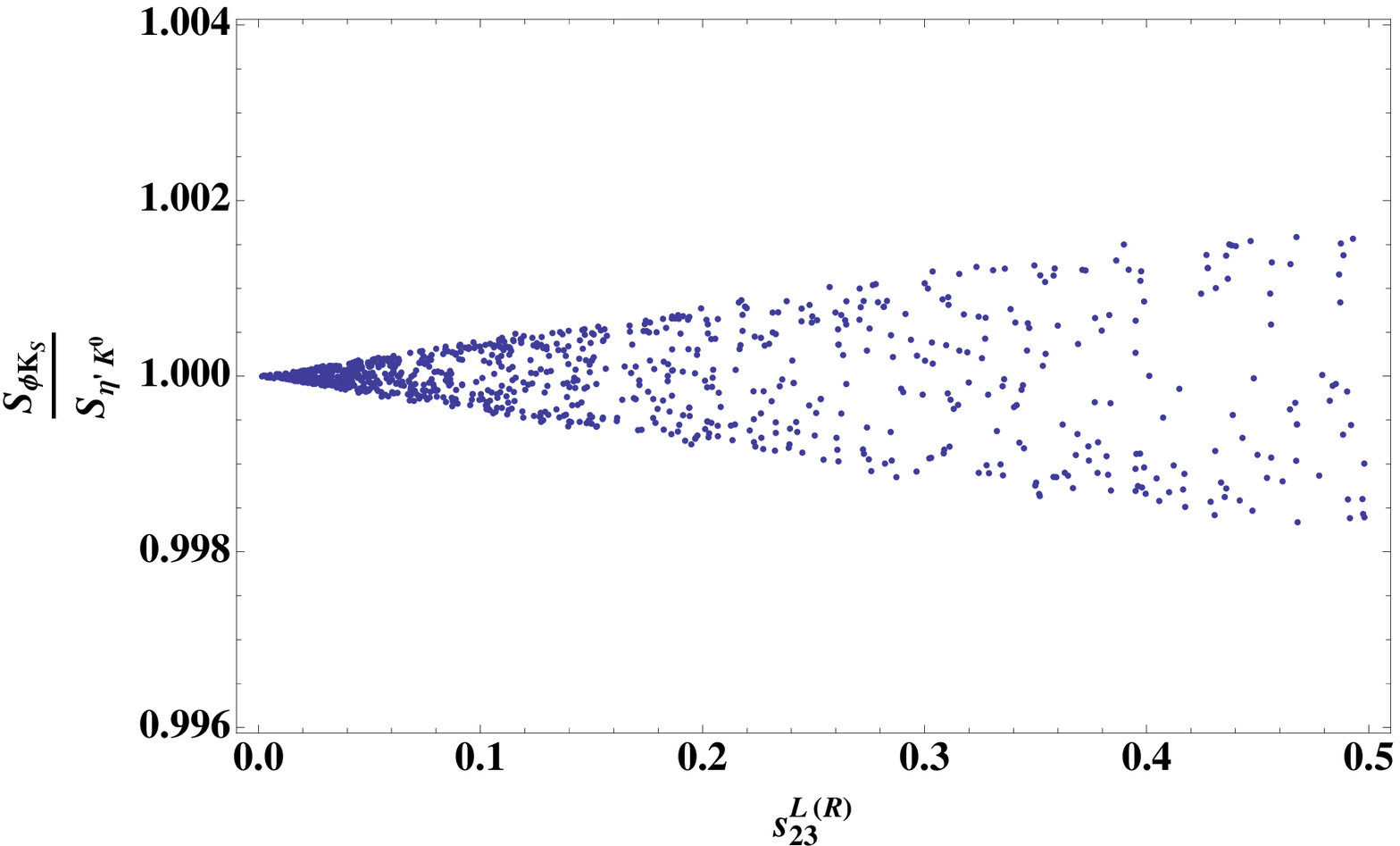}
\hspace{1cm}
\caption{The ratio of $S_{\phi K_S}$ to  $S_{\eta 'K^0}$  versus  $s_{23}^{L(R)}$. 
} 
\label{Sphi}
\end{minipage}
\hspace{1cm}
\begin{minipage}[]{0.45\linewidth}
\includegraphics[width=7.5cm]{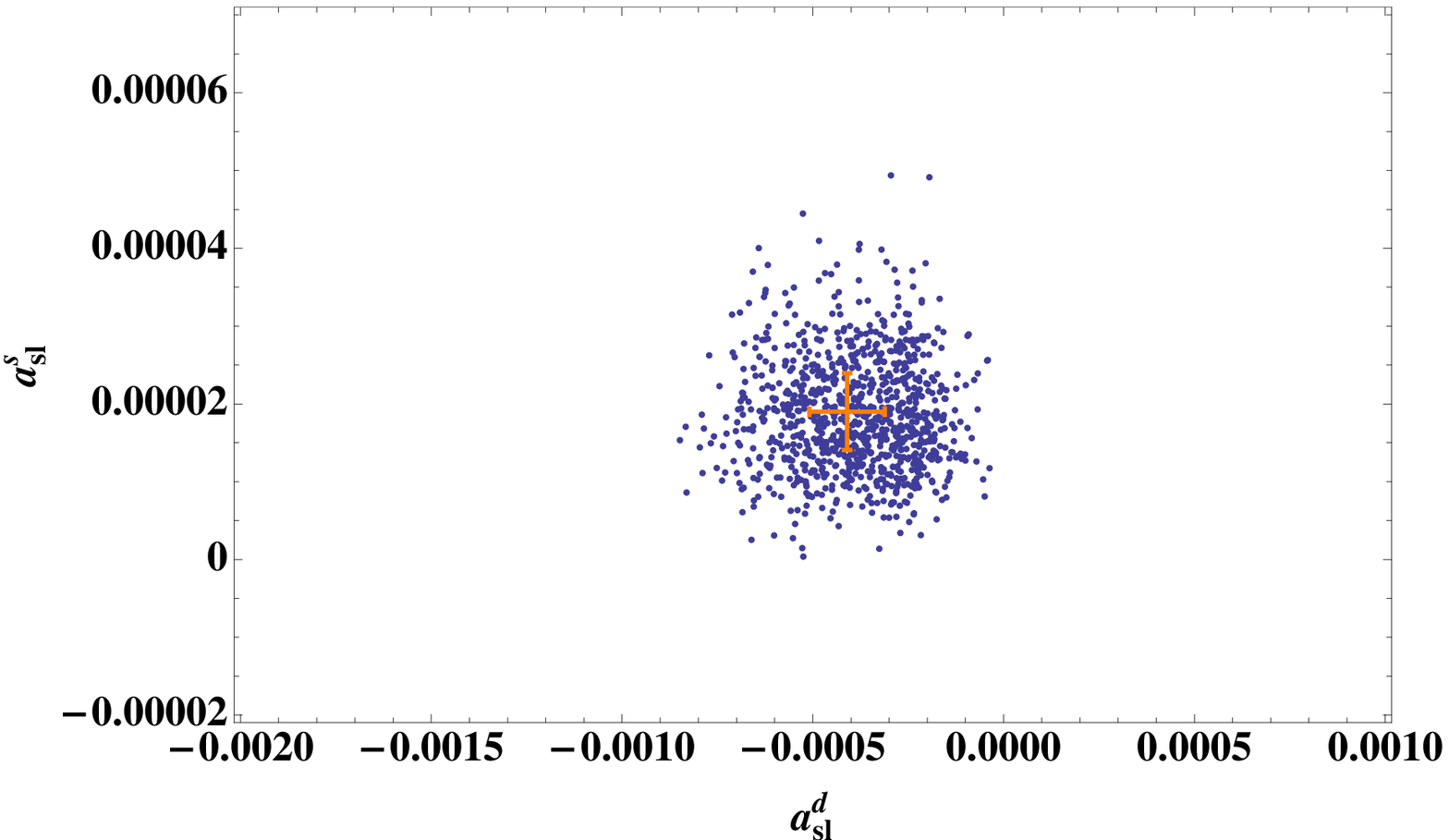}
\hspace{1cm}
\caption{Predicted region of the semileptonic CP asymmetries $a_{sl}^d$ and  $a_{sl}^s$.
The SM prediction is shown by the pink region. } 
\label{semileptonic}
\end{minipage}
\end{figure}
%%%%%%%%%%%%%%%%%%%%%%%%%%%%%%%%%%%%%%%%%%%%%%%%%%%%%%%%%%%
%%%%%%%%%%%%%%%%%%%%%%%%%%%%%%%%%%%%%%%%%%%%%%%%%%%%%%%%%%%
\begin{figure}[!h]
\begin{minipage}[]{0.45\linewidth}
\includegraphics[width=7.8cm]{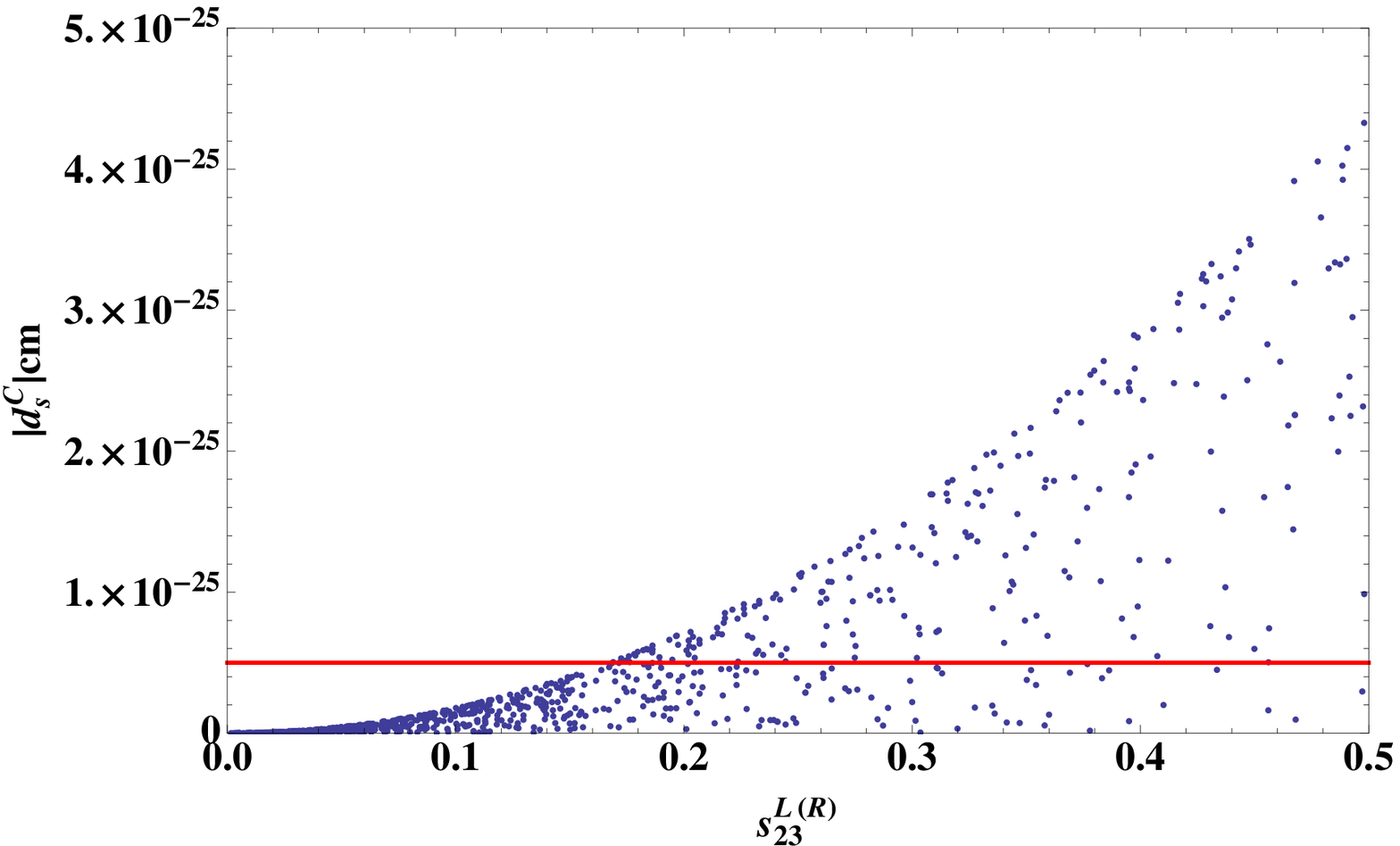}
\hspace{1cm}
\caption{The predicted cEDM of the strange quark versus $s_{23}^L=s_{23}^R$.
The horizontal line denote the experimental upper bound.} 
\label{cEDM}
\end{minipage}
\hspace{1cm}
\begin{minipage}[]{0.45\linewidth}
\includegraphics[width=7.5cm]{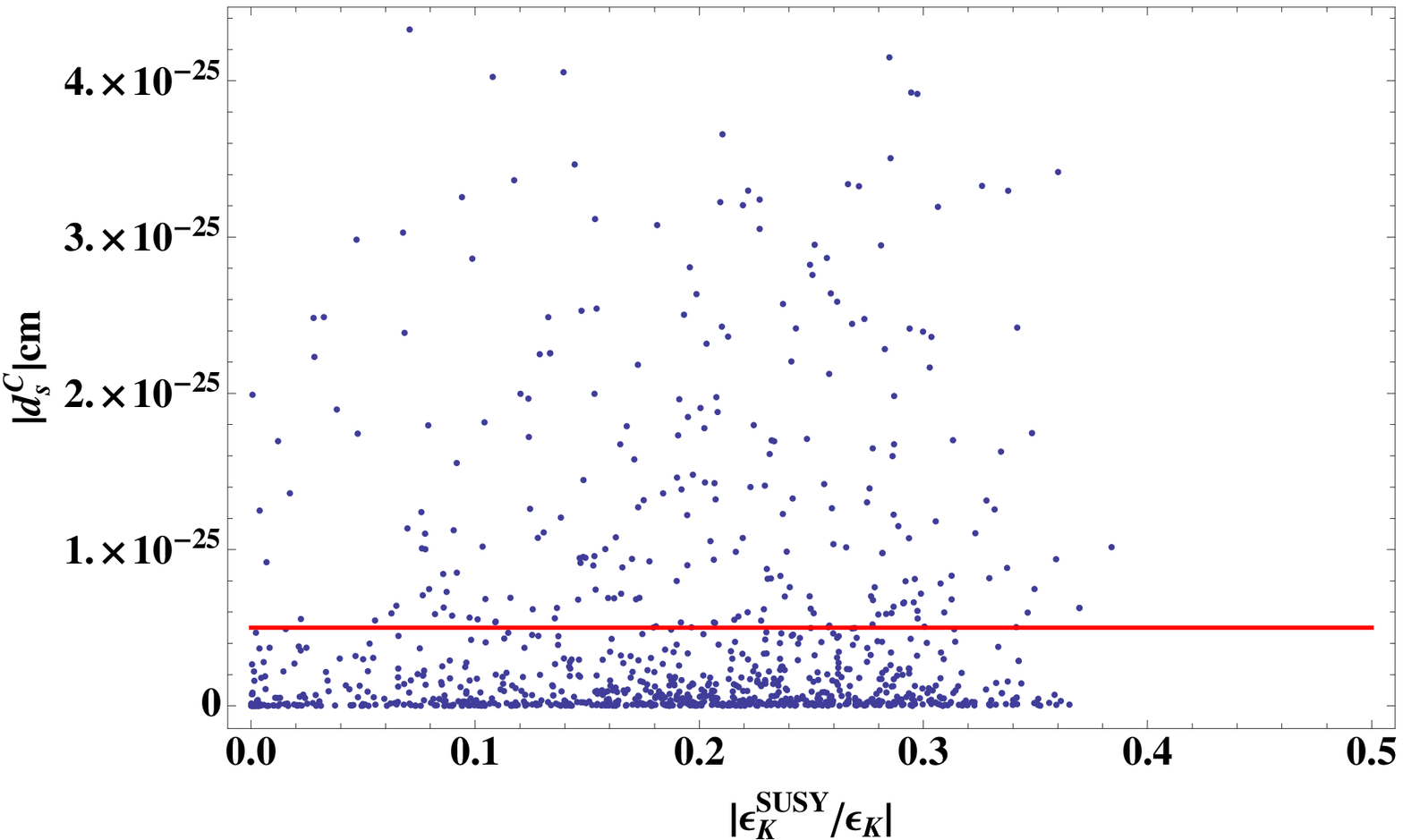}
\hspace{1cm}
\caption{ The predicted cEDM of the strange quark versus $|\epsilon_K^{\rm SUSY} / \epsilon_K|$, where $s_{23}^R= s_{23}^L$. The horizontal line denote the experimental upper bound.} 
\label{epsiloncedm}
\end{minipage}
\end{figure}
%%%%%%%%%%%%%%%%%%%%%%%%%%%%%%%%%%%%%%%%%%%%%%%%%%%%%%%%%%%

Let us discuss the numerical results of  $S_{\phi K_S}$ and $S_{\eta 'K^0}$.
Since $\tilde C_{8G}^{\tilde g}$  is small, the deviation from the SM prediction is also small.
We show the ratio of $S_{\phi K_S}$ to  $S_{\eta 'K^0}$  versus  $s_{23}^{L(R)}$ in Figure \ref{Sphi},
where the SM predicts just one.  The deviation from the SM is tiny, at most $0.2\%$.
Thus,  there is no chance to detect the SUSY contribution in these decay modes.
%We also show the ratio $S_{\phi K_S}$ and $S_{\eta 'K^0}$ versus  $s_{23}^{L(R)}$ in Figure \ref{Sphiratio}.

We discuss  the magnitude of the SUSY contribution to the indirect CP violation 
 $a_{sl}^d$ and $a_{sl}^s$.  We show the predicted  magnitudes  in Figure \ref{semileptonic}.  For the $B^0$ decay, 
the predicted region is 
$a_{sl}^d\simeq -0.001 \sim 0$, on the other hand, for the $B_s$ decay, $a_{sl}^s$ is predicted to be  $a_{sl}^s\simeq  0 \sim 5\times 10^{-5}$, where the SM gives
   $a^{d\rm SM}_{sl}=-(4.1\pm 0.6)\times 10^{-4}$ and 
$a^{s\rm SM}_{sl}=(1.9\pm 0.3)\times 10^{-5}$
as shown in Eq.(\ref{semiCP}).
%We also show  $s_{13}^{L(R)}$  dependence of $a_{sl}^d$ in Figure \ref{semileptonic2}.
%The Belle II experiment is expected to find  the semi-leptonic CP  asymmetry  in ${\cal O}(0.1\%)$.

At the last step, we discuss the cEDM of the strange quark, which depends on  $s_{23}^{L(R)}$.
Under the left-right symmetric assumption $s_{23}^L=s_{23}^R$,
we show the predicted cEDM of the strange quark versus $s_{23}^L(R)$ in Figure \ref{cEDM}.
The predicted cEDM could be larger than the experimental bound of Eq.(\ref{cedm}) , 
$5\times 10^{-26}$cm,
in the region of $s_{23}^{L(R)}\geq 0.17$.

In Figs. 2-11, we have  not imposed the constraint of the  cEDM  of the strange quark.
In order to see the effect of the cEDM constraint, we show the predicted magnitude of  $d_s^C$
 versus $|\epsilon_K^{\rm SUSY}/\epsilon_K|$ in Fig. \ref{epsiloncedm}. Although  some region 
in this plane is excluded by the experimental bound of the cEDM, the allowed region of 
 $|\epsilon_K^{\rm SUSY}/\epsilon_K|$ is not changed.
 This situation is understandable by considering  the different  phase dependence of $\phi^L_{23}$, $\phi^R_{23}$
and $\phi$ for  $d_s^C$ and $\epsilon_K^{\rm SUSY}$, respectively. 
Thus, the  constraint of the cEDM of the strange quark does not change our predictions
although some region of free phase  parameters is excluded.

 In addition, it is noticed that our  result of  $d_s^C$ depends on the assumption $s_{23}^L=s_{23}^R$ considerably.
If we take the suppressed right-handed mixing $s_{23}^R/s_{23}^L=0.1$, the predicted cEDM is just one order reduced, 
on the other hand, $\epsilon_K$ still have $40\%$ contribution of the squark flavor mixing
even in this case.
%In order to see this situation,  we plot the predicted region on the plane 
%$|\epsilon_K^{\rm SUSY} / \epsilon_K|$ and $|d_s^C|$ in Figure \ref{quarkcEDMLL}.

%%%%%%%%%%%%%%%%%%%%%%%%%%%%%%%%%%%%%%%%%%%%%%%%%%%%%%%%%%%
\begin{figure}[t]
\begin{minipage}[]{0.45\linewidth}
\includegraphics[width=7.5cm]{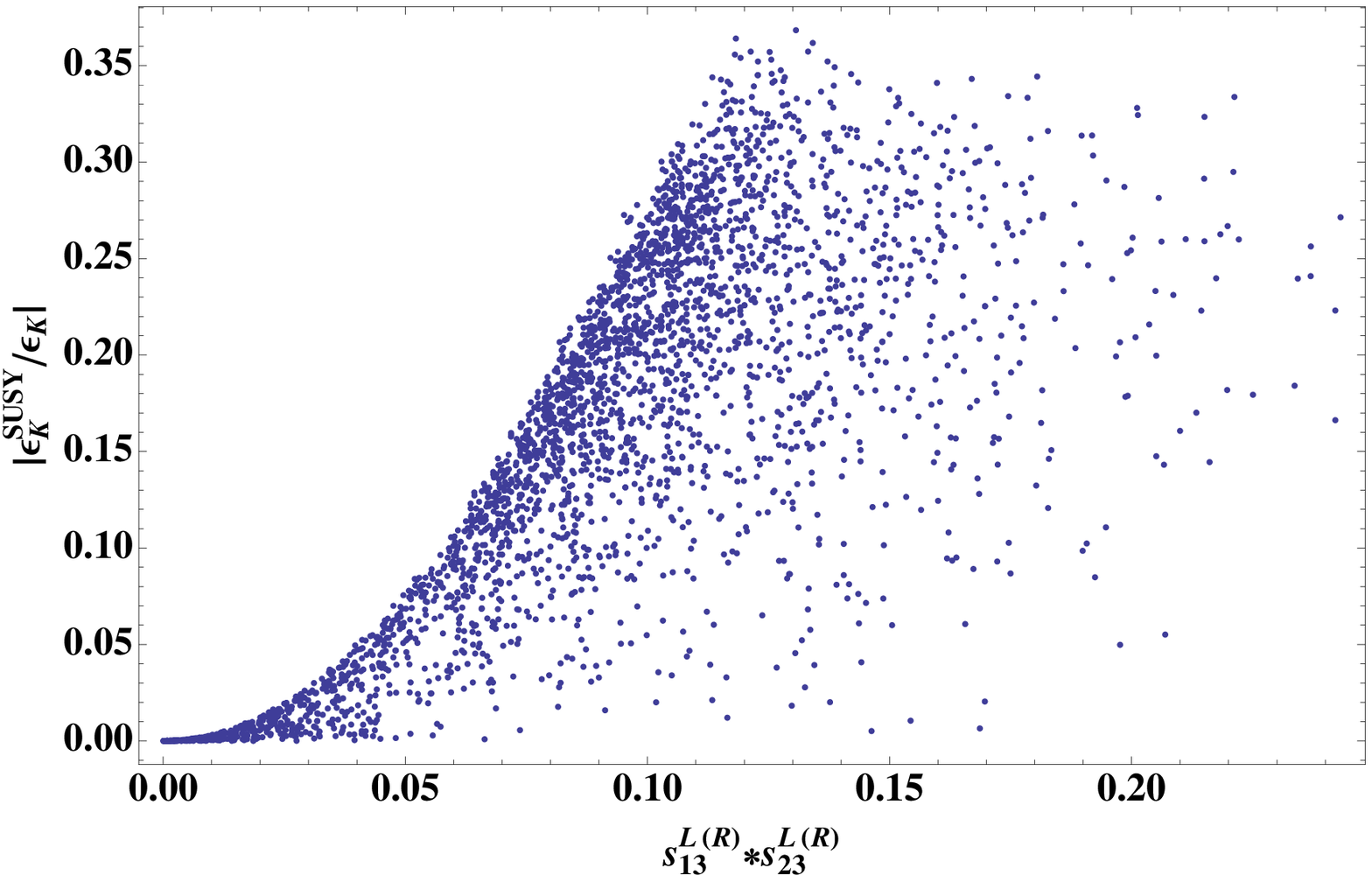}
\hspace{1cm}
\caption{The  $|\epsilon_K^{\rm SUSY} / \epsilon_K|$ versus
$s_{13}^{\rm L(R)} \times s_{23}^{\rm L(R)}$ for $Q_0=50$ TeV.  } 
\label{epsilon50}
\end{minipage}
\hspace{1cm}
\begin{minipage}[]{0.45\linewidth}
\includegraphics[width=7.5cm]{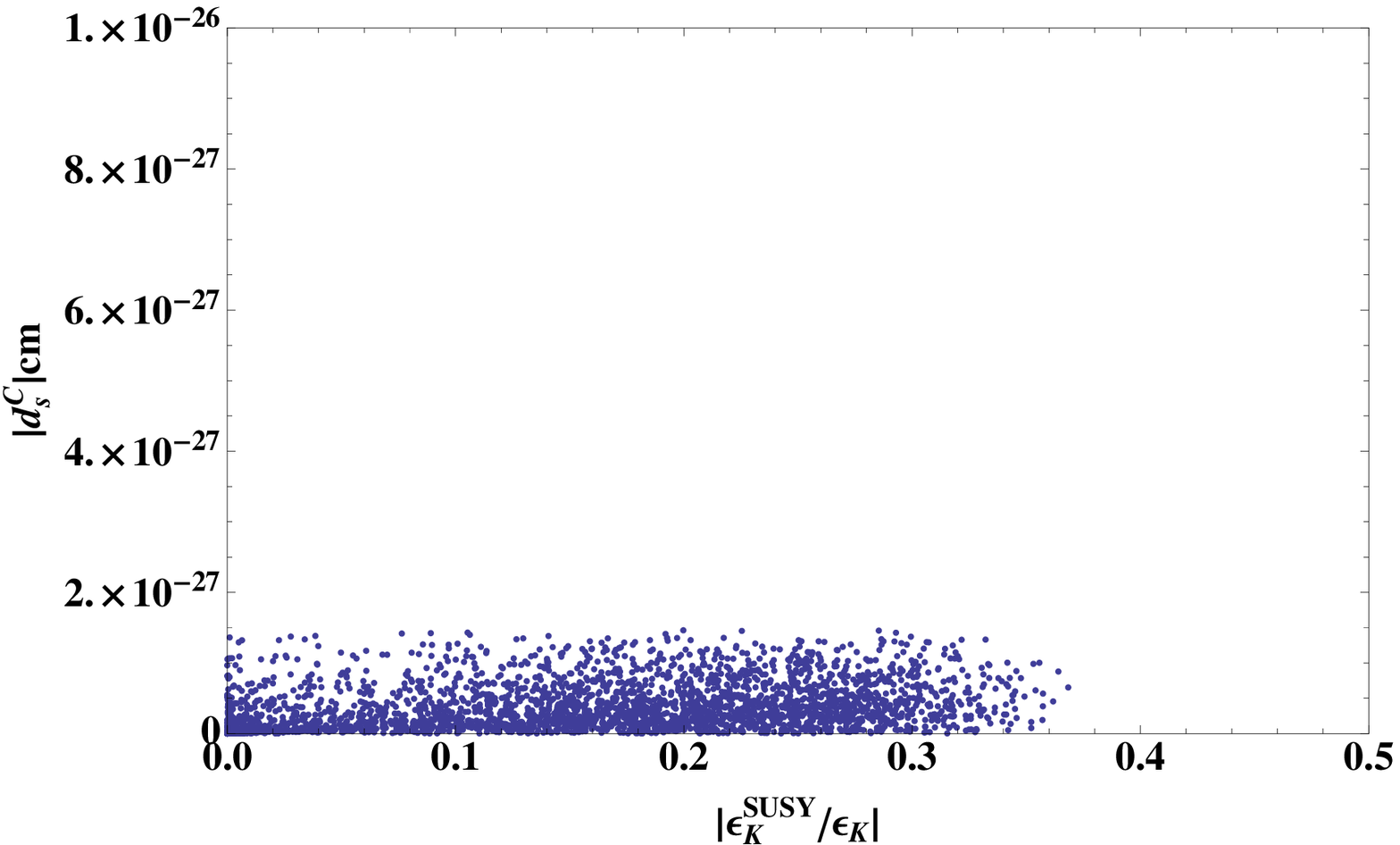}
\hspace{1cm}
\caption{The predicted cEDM versus  $|\epsilon_K^{\rm SUSY} / \epsilon_K|$ 
for $Q_0=50$ TeV. } 
\label{cedm50}
\end{minipage}
\end{figure}
%%%%%%%%%%%%%%%%%%%%%%%%%%%%%%%%%%%%%%%%%%%%%%%%%%%%%%%%%%%

Let us discuss the typical mixing angles of $s_{13}^{L(R)}$ and  $s_{23}^{L(R)}$
in our results.
They are $0.1(0.2)$ for  sizable  SUSY contributions  as seen in  Fig. 3.
These mixing angles are much larger than
the CKM mixing elements  $V_{cb}$ and $V_{ub}$.  
Therefore, 
non-vanishing off diagonal squark mass matrix elements are required  at the $\Lambda$ scale
as discussed below Eq.(\ref{universal}).
For  our squark mass spectrum, the mixing angle  $0.1(0.2)$ corresponds to the off diagonal elements  $(m_{\tilde Q}^2)_{13}$ and $(m_{\tilde Q}^2)_{23}$ to be $\sim 8(16) {\rm TeV}^2$  in the left-handed squark mass matrix.
 Due to the top-Yukawa coupling, the off diagonal element  increases  approximately  $1.4$ times at the $\Lambda$ scale
 compared with the one at the  $Q_0$ scale by the RGE's evolution,
that is  $\sim10(20 ){\rm TeV}^2$ while the diagonal component is  $100 {\rm TeV}^2$.
 Thus,  the universal soft masses should be considered in the  approximation of $10(20)\%$.

%%%%%%%%%%%%%%%%%%%%%%%%%%%%%%%%%%%%%%%%%%%%%%%%%%%%%%%%%%%%%
Let us briefly discuss the case (b) $Q_0=50$ TeV.
The CP violations sensitive to the SUSY contribution is only $\epsilon_K$. 
In the Figure \ref{epsilon50}, we  show the  $|\epsilon_K^{\rm SUSY} / \epsilon_K|$
versus $s_{13}^{\rm L(R)} \times s_{23}^{\rm L(R)}$.
The SUSY contribution could   be also  large up to $35\%$.
Thus,  $\epsilon_K$ is still sensitive to the gluino-squark interaction even if the SUSY scale is $50$ TeV.
This trend continue to the scale $Q_0=100$ TeV.
On the other hand,  cEDM  is reduced to much smaller than the experimental upper bound,
$5\times 10^{-26} $ cm, as seen in Figure \ref{cedm50}.
The situation is different from the one in the case of  $Q_0=10$ TeV.
This result is understandable because the SUSY mass scale increases by five times  and the left-right mixing angle $\theta$
is reduced from $0.35^\circ$ to $0.05^\circ$ compared with the case of $Q_0=10$ TeV as seen in Table 1.

We summarize our results in Table 2, where the sensitivity of the SUSY contribution is presented
for  the case of $Q_0=10$ TeV and $50$ TeV.
 Most sensitive quantity  of the SUSY contribution is  $\epsilon_K$.
 However, more works are required to extract the  SUSY contribution in $\epsilon_K$.
 The unitarity fit is needed to find any mismatch in the SM and single out the SUSY contribution.
 In order to obtain the more precise SM calculation for $\epsilon_K$, 
 the uncertainties of  $\hat B_K$, $V_{cb}$ and $m_t$ must be reduced.
 
The SUSY contributions for  $S_{ J/\psi K_S}$, $S_{ J/\psi \phi}$  and  $\Delta M_{B^0}$ are at most $6-8 \%$.  
Since the theoretical uncertainties in the SM is more than $10\%$, which mainly comes from $\bar\rho$ and $\bar\eta$, 
it is difficult to detect the deviations of $6-8 \%$ from the SM at present.
We hope the precise determination of $\bar\rho$ and $\bar\eta$ in order to find the SUSY
contribution of this level.

As seen in Table 2,  the qualitative features at the $10$ TeV and $50$ TeV scale 
are almost same except for the cEDM of the strange quark.
There is a big chance to observe the neutron EDM in the near future if the SUSY scale is at $10$ TeV.
   
Before closing the presentation of the numerical results, we add a comment on the other gaugino  contribution.
Since left-handed squarks form SU(2) doublets, the mixing angle $\theta_{ij}^L$ also appear 
in the up-type squark mixing matrix. Consequently, there are additional contributions to 
the CP violations of $K$, $B^0$ and $B_s$ mesons induced by chargino exchanging diagrams.
We have obtained  the ratio of the  chargino contribution to the gluino one
for  ${\rm Im} M_{12}(K)$, ${\rm Im} M_{12}^d(B^0)$ and   ${\rm Im} M_{12}^s(B_s)$ as
 $6\%$, $10\%$ and $10\%$, respectively. 
Thus, the  chargino contributions are the sub-leading ones.

%%%%%%%%%%%%%%%%%%%%%%%%%%%%%%%%%%%%%%%%%%%%%%%%%%%%%%%%%
\begin{table}[h]
\begin{center}
\begin{tabular}{|c|c|c|}\hline%|c|l|l|
 & (a) $ Q_0=10$ TeV & (b) $Q_0=50$ TeV \\
\hline
  & & \\
$|\epsilon_K|$ & $ 40\%$ &  $35\%$ \\ 
$S_{ J/\psi K_S}$ &  $6\%$ &  $0.1\%$ \\ 
 $S_{ J/\psi \phi}$  & $ 8\%$ &  $ 0.1\%$ \\ 
 $\Delta M_{B^0}$ & $ 6\%$ &  $0.1\%$ \\ 
  $\Delta M_{B_s}$ & $0.4\%$ &  $0.005\%$ \\ 
 $|S_{\phi K_S}/S_{\eta 'K^0}|-1$ & $0.2\%$ &  $0.001\%$ \\ 
  ${\rm BR}(b\rightarrow s\gamma)$ & $0.3\%$ &  $0.001\%$ \\ 
 $|a^{d}_{sl}|$ & $ \leq 1\times 10^{-3}$ & $\leq 8\times 10^{-4}$ \\ 
$|a^{s}_{sl}|$ & $\leq  5\times 10^{-5}$ &  $\leq 4\times 10^{-5}$ \\ 
$|d^C_s|$&   $ \leq 4\times 10^{-25}$cm &  $\leq 1\times 10^{-27}$cm \\ 
 & & \\
\hline
\end{tabular}
\end{center}
\caption{The  SUSY contribution in the cases (a) $Q_0= 10$ TeV and (b)  $Q_0= 50$ TeV.
 The percents denote  ratios of the SUSY contributions. }
\end{table}
%%%%%%%%%%%%%%%%%%%%%%%%%%%%%%%%%%%%%%%%%%%%%%%%%%%%%%%%%%%

\section{Summary}

We have probed the high scale SUSY, which is at $10$ TeV-$50$ TeV scale, 
in the CP violations of $K$, $B^0$ and $B_s$ mesons.
In order to estimate the contribution of the squark flavor mixing to the CP violations,
we discuss the squark mass spectrum, which is consistent with
the recent Higgs discovery. Taking  the universal soft parameters at the SUSY breaking scale, we obtain
the squark mass spectrum at $10$ TeV and $50$ TeV,  where the SM emerges, by using the RGE's of MSSM. 
And then, the $6\times 6$ mixing matrix between  down-squarks and down-quarks is examined 
by input of the experimental data of $K$, $B^0$ and $B_s$ mesons.
 
It is found that $\epsilon_K$ is most sensitive to the SUSY even if the SUSY scale is at $50$ TeV.
Therefore,  the estimate of  $\epsilon_K$ should be improved 
by reducing uncertainties of the theoretical and experimental input in the SM.
The SUSY contributions for $S_{J/\psi K_S}$, $S_{ J/\psi \phi}$  and  $\Delta M_{B^0}$
are $6-8 \%$ at the SUSY scale of $10$ TeV.
The precise determination of $\bar\rho$ and $\bar\eta$ are required in order to find the SUSY
contribution of this level.
 
We also discussed the high scale SUSY contribution in  the semileptonic CP asymmetry of $B^0$  meson.
We expect the Belle II experiment searching for the semileptonic CP asymmetry $a_{sl}^d$
to find the deviation from the one of the SM in future.
Although  the magnitude of  cEDM of the strange quark depends on  $s_{23}^R/s_{23}^L$ ratio and
the left-right mixing angle of squarks considerably,
there is a big chance to find  the high scale SUSY by the observation of the neutron EDM.
  
In this work, we have discussed  only the down quark-squark sector.
We will study the up quark-squark and lepton-slepton sectors elsewhere.
  
%%%%% acknowledgement %%%%%
\vspace{0.5 cm}
\noindent
{\bf Acknowledgment}

%We thank xxx  for useful discussions. 
This work is supported by JSPS Grand-in-Aid for Scientific Research,
 21340055 and 24654062, 25-5222, respectively.
%%%%%%%%%%%%%%%%%%%%%%%%%%%%%%%%%%%%%%%%%%%%%%%%%%%%%%%%%%%%
%%%%%%%%%%%%%%%%%%%%%%%  Appendices %%%%%%%%%%%%%%%%%%%%%%%%
%%%%%%%%%%%%%%%%%%%%%%%%%%%%%%%%%%%%%%%%%%%%%%%%%%%%%%%%%%%%
\appendix{}
\section*{Appendix}
\section{Squark contribution in $\Delta F=2$ process}

The $\Delta F=2$ effective Lagrangian from the gluino-sbottom-quark interaction  is given as
\begin{align}
\mathcal{L}_{\text{eff}}^{\Delta F=2}=-\frac{1}{2}\left [C_{VLL}O_{VLL}+C_{VRR}O_{VRR}\right ] 
-\frac{1}{2}\sum _{i=1}^2
\left [C_{SLL}^{(i)}O_{SLL}^{(i)}+C_{SRR}^{(i)}O_{SRR}^{(i)}+C_{SLR}^{(i)}O_{SLR}^{(i)}\right ]
\label{Lagrangian-DeltaF=2}
\end{align}
then, the $P^0$-$\bar P^0$ mixing, $M_{12}$, is written as 
\begin{equation}
M_{12}=-\frac{1}{2m_P}\langle P^0|\mathcal{L}_{\text{eff}}^{\Delta F=2}|\bar P^0\rangle \ .
\end{equation}
The hadronic matrix elements are given in terms of the non-perturbative
parameters  $B_i$ as: 
\begin{align}
\langle P^0|\mathcal{O}_{VLL}|\bar P^0\rangle &=\frac{2}{3}m_P^2f_P^2B_1, \quad 
\langle P^0|\mathcal{O}_{VRR}|\bar P^0\rangle =\langle P^0|\mathcal{O}_{VLL}|\bar P^0\rangle ,\nonumber \\
\langle P^0|\mathcal{O}_{SLL}^{(1)}|\bar P^0\rangle &=-\frac{5}{12}m_P^2f_P^2R_PB_2, \quad 
\langle P^0|\mathcal{O}_{SRR}^{(1)}|\bar P^0\rangle =\langle P^0|\mathcal{O}_{SLL}^{(1)}|\bar P^0\rangle ,\nonumber \\
\langle P^0|\mathcal{O}_{SLL}^{(2)}|\bar P^0\rangle &=\frac{1}{12}m_P^2f_P^2R_PB_3, \quad 
\langle P^0|\mathcal{O}_{SRR}^{(2)}|\bar P^0\rangle =\langle P^0|\mathcal{O}_{SLL}^{(2)}|\bar P^0\rangle ,\nonumber \\
\langle P^0|\mathcal{O}_{SLR}^{(1)}|\bar P^0\rangle &=\frac{1}{2}m_P^2f_P^2R_PB_4, \quad 
\langle P^0|\mathcal{O}_{SLR}^{(2)}|\bar P^0\rangle =\frac{1}{6}m_P^2f_P^2R_PB_5,
\end{align}
where 
\begin{equation}
R_P=\left (\frac{m_P}{m_Q+m_q}\right )^2,
\end{equation}
with $(P,Q,q)=(B_d,b,d),~(B_s,b,s),~(K,s,d)$.

The Wilson coefficients for the gluino contribution in Eq.~(\ref{Lagrangian-DeltaF=2}) are written as \cite{GotoNote}

\begin{align}
C_{VLL}(m_{\tilde g})&=\frac{\alpha _s^2}{m_{\tilde g}^2}\sum _{I,J=1}^6
(\lambda _{GLL}^{(d)})_I^{ij}(\lambda _{GLL}^{(d)})_J^{ij}
\left [\frac{11}{18}g_{2[1]}(x_I^{\tilde g},x_J^{\tilde g})
+\frac{2}{9}g_{1[1]}(x_I^{\tilde g},x_J^{\tilde g})\right ],\nonumber \\
C_{VRR}(m_{\tilde g})&=C_{VLL}(m_{\tilde g})(L\leftrightarrow R),\nonumber \\
C_{SRR}^{(1)}(m_{\tilde g})&=\frac{\alpha _s^2}{m_{\tilde g}^2}\sum _{I,J=1}^6
(\lambda _{GLR}^{(d)})_I^{ij}(\lambda _{GLR}^{(d)})_J^{ij}
\frac{17}{9}g_{1[1]}(x_I^{\tilde g},x_J^{\tilde g}),\nonumber \\
C_{SLL}^{(1)}(m_{\tilde g})&=C_{SRR}^{(1)}(m_{\tilde g})(L\leftrightarrow R),\nonumber \\
C_{SRR}^{(2)}(m_{\tilde g})&=\frac{\alpha _s^2}{m_{\tilde g}^2}\sum _{I,J=1}^6
(\lambda _{GLR}^{(d)})_I^{ij}(\lambda _{GLR}^{(d)})_J^{ij}
\left (-\frac{1}{3}\right )g_{1[1]}(x_I^{\tilde g},x_J^{\tilde g}),\nonumber \\
C_{SLL}^{(2)}(m_{\tilde g})&=C_{SRR}^{(2)}(m_{\tilde g})(L\leftrightarrow R),\nonumber 
\end{align}
\begin{align}
C_{SLR}^{(1)}(m_{\tilde g})&=\frac{\alpha _s^2}{m_{\tilde g}^2}\sum _{I,J=1}^6
\Bigg \{ (\lambda _{GLR}^{(d)})_I^{ij}(\lambda _{GRL}^{(d)})_J^{ij}
\left (-\frac{11}{9}\right )g_{2[1]}(x_I^{\tilde g},x_J^{\tilde g}) \nonumber \\
&\hspace{2cm}+(\lambda _{GLL}^{(d)})_I^{ij}(\lambda _{GRR}^{(d)})_J^{ij}
\left [\frac{14}{3}g_{1[1]}(x_I^{\tilde g},x_J^{\tilde g})
-\frac{2}{3}g_{2[1]}(x_I^{\tilde g},x_J^{\tilde g})\right ]\Bigg \} ,\nonumber \\
C_{SLR}^{(2)}(m_{\tilde g})&=\frac{\alpha _s^2}{m_{\tilde g}^2}\sum _{I,J=1}^6
\Bigg \{ (\lambda _{GLR}^{(d)})_I^{ij}(\lambda _{GRL}^{(d)})_J^{ij}
\left (-\frac{5}{3}\right )g_{2[1]}(x_I^{\tilde g},x_J^{\tilde g}) \nonumber \\
&\hspace{2cm}+(\lambda _{GLL}^{(d)})_I^{ij}(\lambda _{GRR}^{(d)})_J^{ij}
\left [\frac{2}{9}g_{1[1]}(x_I^{\tilde g},x_J^{\tilde g})
+\frac{10}{9}g_{2[1]}(x_I^{\tilde g},x_J^{\tilde g})\right ]\Bigg \} ,
\end{align}
where
\begin{align}
(\lambda _{GLL}^{(d)})_K^{ij}&=(\Gamma _{GL}^{(d)\dagger })_i^K(\Gamma _{GL}^{(d)})_K^j~,\quad 
(\lambda _{GRR}^{(d)})_K^{ij}=(\Gamma _{GR}^{(d)\dagger })_i^K(\Gamma _{GR}^{(d)})_K^j~,\nonumber \\
(\lambda _{GLR}^{(d)})_K^{ij}&=(\Gamma _{GL}^{(d)\dagger })_i^K(\Gamma _{GR}^{(d)})_K^j~,\quad 
(\lambda _{GRL}^{(d)})_K^{ij}=(\Gamma _{GR}^{(d)\dagger })_i^K(\Gamma _{GL}^{(d)})_K^j~.
\end{align}
Here we take $(i,j)=(1,3),~(2,3),~(1,2)$ which correspond to $B^0$, $B_s$, and $K^0$ mesons, respectively. 
The loop functions are given as follows:
\begin{itemize}
\item If $x_I^{\tilde g}\not =x_J^{\tilde g}$ ($x_{I,J}^{\tilde g}=m_{\tilde d_{I,J}}^2/m_{\tilde g}^2$),
\begin{align}
g_{1[1]}(x_I^{\tilde g},x_J^{\tilde g})&=\frac{1}{x_I^{\tilde g}-x_J^{\tilde g}}
\left (\frac{x_I^{\tilde g}\log x_I^{\tilde g}}{(x_I^{\tilde g}-1)^2}
-\frac{1}{x_I^{\tilde g}-1}-\frac{x_J^{\tilde g}\log x_J^{\tilde g}}{(x_J^{\tilde g}-1)^2}
+\frac{1}{x_J^{\tilde g}-1}\right ),\nonumber \\
g_{2[1]}(x_I^{\tilde g},x_J^{\tilde g})&=\frac{1}{x_I^{\tilde g}-x_J^{\tilde g}}
\left (\frac{(x_I^{\tilde g})^2\log x_I^{\tilde g}}{(x_I^{\tilde g}-1)^2}
-\frac{1}{x_I^{\tilde g}-1}-\frac{(x_J^{\tilde g})^2\log x_J^{\tilde g}}{(x_J^{\tilde g}-1)^2}
+\frac{1}{x_J^{\tilde g}-1}\right ).
\end{align}
\item If $x_I^{\tilde g}=x_J^{\tilde g}$,
\begin{align}
g_{1[1]}(x_I^{\tilde g},x_I^{\tilde g})&=
-\frac{(x_I^{\tilde g}+1)\log x_I^{\tilde g}}{(x_I^{\tilde g}-1)^3}+\frac{2}{(x_I^{\tilde g}-1)^2}~,\nonumber \\
g_{2[1]}(x_I^{\tilde g},x_I^{\tilde g})&=
-\frac{2x_I^{\tilde g}\log x_I^{\tilde g}}{(x_I^{\tilde g}-1)^3}+\frac{x_I^{\tilde g}+1}{(x_I^{\tilde g}-1)^2}~.
\end{align}
\end{itemize}
 Taking account of the case that  the gluino mass is  much smaller than the squark mass scale $Q_0$,
the effective Wilson coefficients are given at the leading order of QCD as follows:
\begin{align}
C_{VLL}(m_b(\Lambda =2~\text{GeV}))=&\eta _{VLL}^{B(K)}C_{VLL}(Q_0),\quad 
C_{VRR}(m_b(\Lambda =2~\text{GeV}))=\eta _{VRR}^{B(K)}C_{VLL}(Q_0),\nonumber \\
\begin{pmatrix}
C_{SLL}^{(1)}(m_b(\Lambda =2~\text{GeV})) \\
C_{SLL}^{(2)}(m_b(\Lambda =2~\text{GeV}))
\end{pmatrix}&=
\begin{pmatrix}
C_{SLL}^{(1)}(Q_0) \\
C_{SLL}^{(2)}(Q_0)
\end{pmatrix}X_{LL}^{-1}\eta _{LL}^{B(K)}X_{LL},\nonumber \\
\begin{pmatrix}
C_{SRR}^{(1)}(m_b(\Lambda =2~\text{GeV})) \\
C_{SRR}^{(2)}(m_b(\Lambda =2~\text{GeV}))
\end{pmatrix}&=
\begin{pmatrix}
C_{SRR}^{(1)}(Q_0) \\
C_{SRR}^{(2)}(Q_0)
\end{pmatrix}X_{RR}^{-1}\eta _{RR}^{B(K)}X_{RR},\nonumber \\
\begin{pmatrix}
C_{SLR}^{(1)}(m_b(\Lambda =2~\text{GeV})) \\
C_{SLR}^{(2)}(m_b(\Lambda =2~\text{GeV}))
\end{pmatrix}&=
\begin{pmatrix}
C_{SLR}^{(1)}(Q_0) \\
C_{SLR}^{(2)}(Q_0)
\end{pmatrix}X_{LR}^{-1}\eta _{LR}^{B(K)}X_{LR},
\end{align}
where 
\begin{align}
&\eta _{VLL}^B=\eta _{VRR}^B=\left (\frac{\alpha _s({Q_0})}{\alpha _s(\tilde g)}\right )^{\frac{6}{15}}
\left (\frac{\alpha _s(m_{\tilde g})}{\alpha _s(m_t)}\right )^{\frac{6}{21}}
\left (\frac{\alpha _s(m_t)}{\alpha _s(m_b)}\right )^{\frac{6}{23}},\nonumber \\
&\eta _{LL}^B=\eta _{RR}^B=
S_{LL}
\begin{pmatrix}
\eta _{b\tilde g}^{d_{LL}^1} & 0 \\
0 & \eta _{b\tilde g}^{d_{LL}^2}
\end{pmatrix}
S_{LL}^{-1},\qquad 
\eta _{LR}^B=S_{LR}
\begin{pmatrix}
\eta _{b\tilde g}^{d_{LR}^1} & 0 \\
0 & \eta _{b\tilde g}^{d_{LR}^2}
\end{pmatrix}
S_{LR}^{-1},\nonumber \\
&\eta _{b\tilde g}=\left (\frac{\alpha _s({Q_0})}{\alpha _s(m_{\tilde g})}\right )^{\frac{1}{10}}
\left (\frac{\alpha _s(m_{\tilde g})}{\alpha _s(m_t)}\right )^{\frac{1}{14}}
\left (\frac{\alpha _s(m_t)}{\alpha _s(m_b)}\right )^{\frac{3}{46}},\nonumber \\
&\eta _{VLL}^K=\eta _{VRR}^K=\left (\frac{\alpha _s({Q_0})}{\alpha _s (m_{\tilde g})}\right )^{\frac{6}{15}}
\left (\frac{\alpha _s(m_{\tilde g})}{\alpha _s(m_t)}\right )^{\frac{6}{21}}
\left (\frac{\alpha _s(m_t)}{\alpha _s(m_b)}\right )^{\frac{6}{23}}
\left (\frac{\alpha _s(m_b)}{\alpha _s(\Lambda =2~\text{GeV})}\right )^{\frac{6}{25}},\nonumber
\end{align}
\begin{align}
&\eta _{LL}^K=\eta _{RR}^K=
S_{LL}
\begin{pmatrix}
\eta _{\Lambda \tilde g}^{d_{LL}^1} & 0 \\
0 & \eta _{\Lambda \tilde g}^{d_{LL}^2}
\end{pmatrix}
S_{LL}^{-1},\qquad 
\eta _{LR}^K=
S_{LR}
\begin{pmatrix}
\eta _{\Lambda \tilde g}^{d_{LR}^1} & 0 \\
0 & \eta _{\Lambda \tilde g}^{d_{LR}^2}
\end{pmatrix}
S_{LR}^{-1},\nonumber \\
&\eta _{\Lambda \tilde g}=\left (\frac{\alpha _s({Q_0})}{\alpha _s(m_{\tilde g})}\right )^{\frac{1}{10}}
\left (\frac{\alpha _s(m_{\tilde g})}{\alpha _s(m_t)}\right )^{\frac{1}{14}}
\left (\frac{\alpha _s(m_t)}{\alpha _s(m_b)}\right )^{\frac{3}{46}}
\left (\frac{\alpha _s(m_b)}{\alpha _s(\Lambda =2~\text{GeV})}\right )^{\frac{3}{50}},\nonumber \\
&d_{LL}^1=\frac{2}{3}(1-\sqrt{241}),\qquad d_{LL}^2=\frac{2}{3}(1+\sqrt{241}),\qquad 
d_{LR}^1=-16,\qquad d_{LR}^2=2,\nonumber \\
&S_{LL}=
\begin{pmatrix}
\frac{16+\sqrt{241}}{60} & \frac{16-\sqrt{241}}{60} \\
1 & 1
\end{pmatrix},\quad 
S_{LR}=
\begin{pmatrix}
-2 & 1 \\
3 & 0
\end{pmatrix},\nonumber \\
&X_{LL}=X_{RR}=
\begin{pmatrix}
1 & 0 \\
4 & 8
\end{pmatrix},\qquad 
X_{LR}=
\begin{pmatrix}
0 & -2 \\
1 & 0
\end{pmatrix}.\nonumber \\
\end{align}

%%%%%%%%%%%%%%%%%%%%%%%%%%%%%%%%%%%%%%%%%%%%%%%%%
%%%%% Parameters %%%%%%%%%%%%%%%%%%%%%%%%%%%%%%%%
%%%%%%%%%%%%%%%%%%%%%%%%%%%%%%%%%%%%%%%%%%%%%%%%%

For the parameters $B_i^{(d)}(i=2-5)$ of $B$ mesons,  we use values in  \cite{Becirevic:2001xt}
as follows:
\begin{eqnarray}
&&B_2^{(B_d)} (m_b)=0.79(2)(4), \qquad
B_3^{(B_d)} (m_b)=0.92(2)(4), \nonumber \\
&&B_4^{(B_d)} (m_b)=1.15(3)(^{+5}_{-7}), \qquad 
B_5^{(B_d)} (m_b)=1.72(4)(^{+20}_{-6}), \nonumber\\
&&B_2^{(B_s)} (m_b)=0.80(1)(4), \qquad
B_3^{(B_s)} (m_b)=0.93(3)(8), \nonumber\\
&&B_4^{(B_s)} (m_b)=1.16(2)(^{+5}_{-7}), \qquad
B_5^{(B_s)} (m_b)=1.75(3)(^{+21}_{-6})\ .
\end{eqnarray}
On the other hand, we use the most updated values for $\hat B_1^{(d)} $ and
 $\hat B_1^{(s)} $ as \cite{UTfit}
\begin{equation}
\hat B_1^{(B_s)}  = 1.33\pm 0.06 \ , \qquad  %KEKFFlattice result  Marco Ciuchini \\
\hat B_1^{(B_s)} / \hat B_1^{(B_d)}=1.05\pm 0.07 \ . % KEKFFlattice result \\
\end {equation}

For the paremeters $B_i^{K}(i=2-5)$, we use following values 
  \cite{Allton:1998sm},
\begin{equation}
\begin{split}
B_2^{(K)}(2{\rm GeV})=0.66\pm 0.04 , \qquad 
B_3^{(K)}(2{\rm GeV})=1.05\pm 0.12, \\
B_4^{(K)}(2{\rm GeV})=1.03\pm 0.06, \qquad
B_5^{(K)}(2{\rm GeV})=0.73\pm 0.10,
\end{split}
\end{equation}
and we take recent value of Eq.(\ref{BK}) for deriving $B_1^{(K)}(2{\rm GeV})$.

\section{The loop functions $F_i$}

The loop functions $F_i(x_{\tilde g}^I)$ are given in terms of  $x_{\tilde g}^I=m_{\tilde g}^2/m_{\tilde d_I}^2~(I=3,6)$
 as
\begin{align}
F_1(x_{\tilde g}^I)&=\frac{x_{\tilde g}^I\log x_{\tilde g}^I}{2(x_{\tilde g}^I-1)^4}
+\frac{(x_{\tilde g}^I)^2-5x_{\tilde g}^I-2}{12(x_{\tilde g}^I-1)^3}~, \quad
F_2(x_{\tilde g}^I)=-\frac{(x_{\tilde g}^I)^2\log x_{\tilde g}^I}{2(x_{\tilde g}^I-1)^4}
+\frac{2(x_{\tilde g}^I)^2+5x_{\tilde g}^I-1}{12(x_{\tilde g}^I-1)^3}~,\nonumber \\
F_3(x_{\tilde g}^I)&=\frac{\log x_{\tilde g}^I}{(x_{\tilde g}^I-1)^3}
+\frac{x_{\tilde g}^I-3}{2(x_{\tilde g}^I-1)^2}~, \quad
F_4(x_{\tilde g}^I)=-\frac{x_{\tilde g}^I\log x_{\tilde g}^I}{(x_{\tilde g}^I-1)^3}+
\frac{x_{\tilde g}^I+1}{2(x_{\tilde g}^I-1)^2}=\frac{1}{2}g_{2[1]}(x_{\tilde g}^I,x_{\tilde g}^I)~,
\end{align}

%Other  Wilson coefficients are omitted in this appendix because those are
% next-leading contributions to our numerical calculations.
%The Wilson coefficients $\widetilde C_i^{\tilde g}(m_{\tilde g})$'s are 
%obtained by replacing $L(R)$ with $R(L)$ in $C_i^{\tilde g}(m_{\tilde g})$'s. 

\section{cEDM}
The cEDM of the strange quark from gluino contribution is given by \cite{GotoNote}
\begin{equation}
d_s^C(Q_0)=-2\sqrt{4\pi \alpha _s(m_{\tilde g})}\text{Im}[A_s^{g22}(Q_0)],
\end{equation}
where 
\begin{align}
%A_s^{g22}&=-\frac{\alpha _s(m_{\tilde g})}{24\pi }
A_s^{g22}&(Q_0)
=-\frac{\alpha _s(m_{\tilde g})}{4\pi }\frac{1}{3}
\Bigg [\frac{1}{2m_{\tilde d_3}^2}\bigg \{ 
\Big (m_s(\lambda _{GLL}^{(d)})_3^{22}+m_s(\lambda _{GRR}^{(d)})_3^{22}\Big )
\Big (9F_1(x_{\tilde g}^3)+F_2(x_{\tilde g}^3)\Big )\nonumber \\
&\hspace{3.3cm}+m_{\tilde g}(\lambda _{GLR}^{(d)})_3^{22}
\Big (9F_3(x_{\tilde g}^3)+F_4(x_{\tilde g}^3)\Big )\bigg \} \\
&\hspace{-0.7cm}+\frac{1}{2m_{\tilde d_6}^2}\bigg \{ 
\Big (m_s(\lambda _{GLL}^{(d)})_6^{22}+m_s(\lambda _{GRR}^{(d)})_6^{22}\Big )
\Big (9F_1(x_{\tilde g}^6)+F_2(x_{\tilde g}^6)\Big )
+m_{\tilde g}(\lambda _{GLR}^{(d)})_6^{22}
\Big (9F_3(x_{\tilde g}^6)+F_4(x_{\tilde g}^6)\Big )\bigg \} \Bigg ].\nonumber
\end{align}

Including the QCD correction, we get
 
 \begin{equation}
d_s^C(2{\rm GeV})=d_s^C(Q_0)
\left ( \frac{\alpha_s(Q_0)}{\alpha_s(m_{\tilde g})} \right )^{\frac{14}{15}}
\left ( \frac{\alpha_s(m_{\tilde g})}{\alpha_s(m_t)} \right )^{\frac{14}{21}}
 \left ( \frac{\alpha_s(m_t)}{\alpha_s(m_b)} \right )^{\frac{14}{23}}
\left ( \frac{\alpha_s(m_b)}{\alpha_s(2{\rm GeV})} \right )^{\frac{14}{25}} \ .
\end{equation}

%\newpage

%%%%%%%%%%%%%%%%%%%%%%%%%%%%%%%%%%%%%%%%%%%%%%%
%%%%%%%%  Regerences %%%%%%%%%%%%%%%%%%%%%%%%%%
%%%%%%%%%%%%%%%%%%%%%%%%%%%%%%%%%%%%%%%%%%%%%%%

\end{document}